\documentclass{elsart}
\usepackage{amssymb}
\usepackage{graphicx}
\usepackage{graphics}
\usepackage{epsfig}
\newtheorem{pp}{Proposition}
\journal{Physica D}

\begin{document}

\begin{frontmatter}

\title{Geometrical properties of local dynamics in Hamiltonian systems:
the Generalized Alignment Index (GALI) method}

\author[Patra,Paris]{Ch. Skokos\corauthref{COR}},
\corauth[COR]{Corresponding author.}
\ead{hskokos@imcce.fr}
\ead[url]{http://www.imcce.fr/$\,\tilde{}\,$hskokos}
\author[Patra]{T.C. Bountis},
\ead{bountis@math.upatras.gr}
\ead[url]{http://www.math.upatras.gr/$\,\tilde{}\,$bountis}
\author[Patra]{Ch. Antonopoulos}
\ead{antonop@math.upatras.gr}
\ead[url]{http://www.math.upatras.gr/$\,\tilde{}\,$antonop}

\address[Patra]{Department of Mathematics, Division of Applied
Analysis and Center for Research and Applications of Nonlinear
Systems (CRANS), University of Patras, GR-26500 Patras, Greece}
\address[Paris]{Astronomie et Syst\`{e}mes Dynamiques, IMCCE,
Observatoire de Paris, 77 avenue Denfert--Rochereau, F-75014,
Paris, France}

\begin{abstract}
We investigate the detailed dynamics of multidimensional
Hamiltonian systems by studying the evolution of volume elements
formed by unit deviation vectors about their orbits. The behavior
of these volumes is strongly influenced by the regular or chaotic
nature of the motion, the number of deviation vectors, their
linear (in)dependence and the spectrum of Lyapunov exponents. The
different time evolution of these volumes can be used to identify
rapidly and efficiently the nature of the dynamics, leading to the
introduction of quantities that clearly distinguish between
chaotic behavior and quasiperiodic motion on $N$-dimensional tori.
More specifically we introduce the Generalized Alignment Index of
order $k$ (GALI$_k$) as the volume of a generalized
parallelepiped, whose edges are $k$ initially linearly independent
unit deviation vectors from the studied orbit whose magnitude is
normalized to unity at every time step. We show analytically and
verify numerically on particular examples of $N$ degree of freedom
Hamiltonian systems that, for chaotic orbits, GALI$_k$ tends
exponentially to zero with exponents that involve the values of
several Lyapunov exponents. In the case of regular orbits,
GALI$_k$ fluctuates around non--zero values for $2\leq k \leq N$
and goes to zero for $N < k \leq 2N$ following power laws that
depend on the dimension of the torus and the number $m$ of
deviation vectors initially tangent to the torus: $\propto
t^{-2(k-N)+m}$ if  $0\leq m <k-N$, and $\propto t^{-(k-N)}$ if $m
\geq k-N$. The GALI$_k$ is a generalization of the Smaller
Alignment Index (SALI) (GALI$_2$ $\propto$ SALI). However,
GALI$_k$ provides significantly more detailed information on the
local dynamics, allows for a faster and clearer distinction
between order and chaos than SALI and works even in cases where
the SALI method is inconclusive.

\end{abstract}

\end{frontmatter}

\section{Introduction}
\label{intro}

Determining the chaotic or regular nature of orbits in {\it
conservative} dynamical systems is a fundamental issue of
nonlinear science. The difficulty with conservative systems, of
course, is that regular and chaotic orbits are distributed
throughout phase space in very complicated ways, in contrast with
dissipative systems, where all orbits eventually fall on regular
or chaotic attractors. Over the years, several methods
distinguishing regular from chaotic motion in conservative systems
have been proposed and applied, with varying degrees of success.
These methods can be divided in two major categories: Some are
based on the study of the evolution of small deviation vectors
from a given orbit, while others rely on the analysis of the
particular orbit itself.

The most commonly employed method for distinguishing between order
and chaos, which belongs to the category related to the study
of deviation vectors, is the evaluation of the maximal Lyapunov
Characteristic Exponent (LCE) $\sigma_1$; if $\sigma_1 > 0$ the
orbit is chaotic. The theory of Lyapunov exponents was applied to
characterize chaotic orbits by Oseledec \cite{O68}, while the
connection between Lyapunov exponents and exponential divergence
of nearby orbits was given in \cite{BGS76,P77}. Benettin et al.\
\cite{BGGS80a} studied the problem of the computation of all LCEs
theoretically and proposed in \cite{BGGS80b} an algorithm for
their numerical computation. In particular, $\sigma_1$ is computed
as the limit for $t \rightarrow \infty$ of the quantity
\begin{equation}
L_1(t)=\frac{1}{t}\, \ln  \frac{\|\vec{w}(t)\|}{\|\vec{w}(0)\|}\,
,\, \mbox{i.e.}\,\, \sigma_1 = \lim_{t\rightarrow \infty} L_1 (t)
\, , \label{eq:lyap1_def}
\end{equation}
where $\vec{w}(0)$, $\vec{w}(t)$ are  deviation vectors from a given
orbit, at times $t=0$ and $t>0$ respectively. It has been shown that
the above limit is finite, independent of the choice of the metric
for the phase space and converges to $\sigma_1$ for almost all
initial vectors $\vec{w}(0)$ \cite{O68,BGGS80a,BGGS80b}. Similarly,
all other LCEs, $\sigma_2$, $\sigma_3$ etc.\ are computed as the
limits for $t \rightarrow \infty$ of some appropriate quantities,
$L_2(t)$, $L_3(t)$ etc.~(see \cite{BGGS80b} for more details). We
note that throughout the present paper, whenever we need to compute
the values of the maximal LCE or of several LCEs
we apply respectively the algorithms proposed by
Benettin et al.\ \cite{BGS76,BGGS80b}. Since 1980, new methods have
been introduced for the effective computation of LCEs (e.~g.\
\cite{GK87}, see also \cite{BR01} and references therein). The true
power of these techniques is revealed in the study of
multi--dimensional systems, when only a small number of LCE are of
interest. In such cases, these methods are significantly more
efficient than the method of \cite{BGGS80b}, which computes the whole
spectrum of LCEs. On the other hand, they are less or equally
efficient when compared with the method of \cite{BGS76} for the
computation of the maximal LCE, whose value is sufficient for the
determination of the regular or chaotic nature of an orbit.

Among other chaoticity detectors, belonging to the same category
with the evaluation of the maximal LCE, are the fast Lyapunov
indicator (FLI) and its variants
\cite{FLG97,FGL97,FLFF02,GLF02,B05}, the mean exponential growth of
nearby orbits (MEGNO) \cite{CS00,CGS03}, the smaller alignment index
(SALI) \cite{S01,SABV03b,SABV04}, the relative Lyapunov indicator
(RLI) \cite{SESF04}, as well as methods based on the study of power
spectra of deviation vectors \cite{VVT00}, as well as spectra of
quantities related to these vectors \cite{FFL93,LFD93,VC94}. In the
category of methods based on the analysis of a time series
constructed by the coordinates of the orbit under study, one may
list the frequency map analysis of Laskar
\cite{L90,LFC92,L93,PL96,PL98,L99}, the method of the low frequency
power (LFP) \cite{VI92,KV04}, the `0--1' test \cite{GM04}, as well
as some other more recently introduced techniques \cite{S05,H05}.

In the present paper, we generalize and improve considerably the
SALI method mentioned above by introducing the Generalized
ALignment Index (GALI). This index retains the advantages of the
SALI -- i.e.~its simplicity and efficiency in distinguishing
between regular and chaotic motion -- but, in addition, is faster
than the SALI, displays power law decays that depend on torus
dimensionality and can also be applied successfully to cases where
the SALI is inconclusive, like in the case of chaotic orbits whose
two largest Lyapunov exponents are equal or almost equal.

For the computation of the GALI we use information from the
evolution of {\it more than two} deviation vectors from the
reference orbit, while SALI's computation requires the evolution
of only two such vectors. In particular, GALI$_k$ is proportional
to `volume' elements formed by $k$ initially linearly independent
unit deviation vectors whose magnitude is normalized to unity at
every time step. If the orbit is chaotic, GALI$_k$ goes to zero
exponentially fast by the law
\begin{equation}
\mbox{GALI}_k(t)  \propto e^{-\left[ (\sigma_1-\sigma_2) +
(\sigma_1-\sigma_3)+ \cdots+
(\sigma_1-\sigma_k)\right]t}. \nonumber\\
\end{equation}
If, on the other hand, the orbit lies in an $N$--dimensional
torus, GALI$_k$ displays the following behaviors: Either
\begin{equation}
\mbox{GALI}_k (t) \approx \mbox{constant} \,\,\, \mbox{for} \,\,\,
2 \leq k \leq N,
\end{equation}
or, if $N<k\leq 2N$, it decays with {\it different power laws},
depending on the number $m$ of deviation vectors which initially
lie in the tangent space of the torus, i.~e.~:
\begin{equation}
\mbox{GALI}_k (t) \propto \left\{ \begin{array}{ll}
\frac{1}{t^{2(k-N)-m}} & \mbox{if $N< k \leq 2N$ and $0\leq m <
k-N$} \\
\frac{1}{t^{k-N}} & \mbox{if $N< k \leq 2N$ and $m \geq
k-N$} \\
\end{array} \right.
\end{equation}

So, the GALI allows us to study more efficiently the {\it
geometrical} properties of the dynamics in the neighborhood of an
orbit, especially in higher dimensions, where it allows for a much
faster determination of its chaotic nature, overcoming the
limitations of the SALI method. In the case of regular motion,
GALI$_k$ is either a constant, or decays by power laws that depend
on the dimensionality of the subspace in which the orbit lies, which
can prove useful e.g., if our orbits are in a `sticky' region, or if
our system happens to possess fewer or more than $N$ independent
integrals of the motion (i.e. is partially integrable or
super-integrable respectively).

This paper is organized as follows: In section \ref{SALI}, we
recall the definition of the SALI describing also its behavior for
regular and chaotic orbits of Hamiltonian flows and symplectic
maps. In section \ref{GALI_d}, we introduce the GALI$_k$ for k
deviation vectors, explaining in detail its numerical computation,
while in section \ref{GALI_b} we study theoretically the behavior
of the new index for chaotic and regular orbits.  Section
\ref{GALI_a} presents applications of the GALI$_k$ approach to
various Hamiltonian systems of different numbers of degrees of
freedom, concentrating on its particular advantages. Finally, in
section \ref{Summary}, we summarize the results and present our
conclusions, while the appendices are devoted respectively to the
definition of the wedge product and the explanation of the
explicit connection between GALI$_2$ and SALI.

\section{The SALI}
\label{SALI}

The SALI method was introduced in \cite{S01} and has  been applied
successfully to detect regular and chaotic motion in Hamiltonian
flows as well as symplectic maps
\cite{SABV03a,SABV03b,S03,SESS04,SABV04,PBS04,AHHN05,ABS05,MA05b,MA05a,MA06,BS05,CDLMV06}.
It is an index that tends exponentially to zero in the case of
chaotic orbits, while it fluctuates around non--zero values for
regular trajectories of Hamiltonian systems and 2$N$--dimensional
symplectic maps with $N>1$. In the case of 2--dimensional (2D)
maps, the SALI tends to zero both for regular and chaotic orbits
but with very different time rates, which allows us again to
distinguish between the two cases \cite{S01}: In particular the
SALI tends to zero following an exponential law for chaotic orbits
and decays to zero following a power law for regular orbits.

The basic idea behind the success of the SALI method (which
essentially distinguishes it from the computation of LCEs) is the
introduction of one additional deviation vector with respect to a
reference orbit. Indeed, by considering the relation between two
deviation vectors (instead of one deviation vector and the
reference orbit), one is able to circumvent the difficulty of the
slow convergence of Lyapunov exponents to non--zero (or zero)
values as $t\rightarrow \infty$.

In order to compute the SALI, therefore, one follows simultaneously
the time evolution of a reference orbit along with two deviation
vectors with initial conditions $\vec{w}_1 (0)$, $\vec{w}_2 (0)$.
Since we are only interested in the directions of these two vectors
we normalize them, at every time step, keeping their norm equal to
1, setting
\begin{equation}
\hat{w}_i (t) =\frac{\vec{w}_i (t)}{\|\vec{w}_i(t)\|}, \, \,\,
i=1,2 \label{eq:w_hat}
\end{equation}
where $\|\cdot\|$ denotes the Euclidean norm and the hat ($^\wedge
$) over a vector denotes that it is of unit magnitude. The SALI is
then defined as:
\begin{equation}
\mbox{SALI}(t)= \min \left\{ \left\| \hat{w}_1 (t)+\hat{w}_2 (t)
\right\| , \left\| \hat{w}_1 (t)-\hat{w}_2 (t) \right\| \right\},
\label{eq:SALI}
\end{equation}
whence it is evident that $\mbox{SALI}(t) \in [0,\sqrt{2}]$.
$\mbox{SALI}=0$ indicates that the two deviation vectors have
become aligned in the same direction (and are equal or opposite to
each other); in other words, they are linearly dependent.

Let us observe, at this point, that seeking the minimum of the two
positive quantities in (\ref{eq:SALI}) (which are bounded above by
2) is essentially equivalent to evaluating the product
\begin{equation}
P(t)= \left\| \hat{w}_1 (t)+\hat{w}_2 (t) \right\| \cdot \left\|
\hat{w}_1 (t)-\hat{w}_2 (t) \right\| , \label{eq:product}
\end{equation}
at every value of $t$. Indeed, if the minimum of these two
quantities is zero (as in the case of a chaotic reference orbit,
see below), so will be the value of $P(t)$. On the other hand, if
it is not zero, $P(t)$ will be proportional to the constant about
which this minimum oscillates (as in the case of regular motion,
see below). This suggests that, instead of computing the SALI$(t)$
from (\ref{eq:SALI}), one might as well evaluate the `exterior' or
`wedge' product of the two deviation vectors $\hat{w}_1 \wedge
\hat{w}_2$ for which it holds
\begin{equation}
\| \hat{w}_1 \wedge \hat{w}_2 \|=
\frac{\|\hat{w}_1-\hat{w}_2\|\cdot
\|\hat{w}_1+\hat{w}_2\|}{2}\,\,\, ,\label{eq:SALI_wedge}
\end{equation}
and which represents the `area' of the parallelogram formed by the
two deviation vectors. For the definition of the wedge product see
Appendix \ref{Wedge} and for a proof of (\ref{eq:SALI_wedge}) see Appendix
\ref{GALI_2}. Indeed, the `wedge' product can provide much more
useful information, as it can be generalized to represent the
`volume' of a parallelepiped formed by the vectors $\hat{w}_1,
\hat{w}_2, \ldots, \hat{w}_k$, $2\leq k \leq 2N$, regarded as
deviations from an orbit of an $N$--degree of freedom Hamiltonian
system, or a $2N$--dimensional symplectic map.

It is the main purpose of this paper to study precisely such a
generalization and reveal considerably more qualitative and
quantitative information about the local and global dynamics of
these systems. Before we proceed to describe this generalization,
however, let us first summarize what we know about the properties
of the SALI  for the case of two deviation vectors $\hat{w}_1$,
$\hat{w}_2$:
\begin{enumerate}
    \item In the case of chaotic orbits, the deviation vectors
$\hat{w}_1$, $\hat{w}_2$ eventually become aligned in the
direction of the maximal Lyapunov exponent, and SALI$(t)$ falls
exponentially to zero. An analytical study of SALI's behavior for
chaotic orbits was carried out in \cite{SABV04} where it was shown
that
\begin{equation}
\mbox{SALI}(t) \propto e^{-(\sigma_1-\sigma_2)t} \label{eq:exp}
\end{equation}
$\sigma_1$, $\sigma_2$ being the two largest LCEs.
   \item  In the case of regular motion, on the other hand, the orbit
lies on a torus and the vectors $\hat{w}_1$, $\hat{w}_2$
eventually fall on  its tangent space, following a $t^{-1}$ time
evolution, having in general different directions. In this case,
the SALI oscillates about values that are different from zero (for
more details see \cite{SABV03b}). This behavior is due to the fact
that for regular orbits the norm of a deviation vector increases
linearly in time along the flow. Thus, our normalization procedure
brings about a decrease of the magnitude of the coordinates
perpendicular to the torus at a rate proportional to $t^{-1}$ and
so $\hat{w}_1$, $\hat{w}_2$ eventually fall on the tangent space
of the torus.
\end{enumerate}

Note that in the case of 2D maps the torus is actually an
invariant curve and its tangent space is 1--dimensional. So, in
this case, the two unit deviation vectors eventually become
linearly dependent and SALI becomes zero following a power law.
This is, of course, different than the exponential decay of SALI
for chaotic orbits and thus SALI can distinguish easily between
the two cases even in 2D maps \cite{S01}. Thus, although the
behavior of SALI in 2D maps is clearly understood, the fact
remains that SALI does not always have the same behavior for
regular orbits, as it may oscillate about a constant or decay to
zero by a power law, depending on the dimensionality of the
tangent space of the reference orbit. It might, therefore, be
interesting to ask whether this index can be generalized, so that
different power laws may be found to characterize regular motion
in higher dimensions. It is one of the principal aims of this
paper to show that such a generalization is possible.

Let us make one final remark concerning the behavior of SALI for
chaotic orbits: Looking at equation (\ref{eq:exp}), one might
wonder what would happen in the case of a chaotic orbit whose two
largest Lyapunov exponents $\sigma_1$ and $\sigma_2$ are equal or
almost equal. Although this may not be common in generic
Hamiltonian systems, such cases can be found in the literature. In
one such example \cite{ABS05}, very close to a particular unstable
periodic orbit of a 15 degree of freedom Hamiltonian system, the
two largest Lyapunov exponents are nearly equal $\sigma_1 -
\sigma_2 \approx 0.0002$. Even though, in that example, SALI still
tends to zero at the rate indicated by (\ref{eq:exp}), it is
evident that the chaotic nature of an orbit cannot be revealed
very fast by the SALI method. It is, therefore, clear that a more
detailed analysis of the local dynamics is needed to further
explore the properties of specific orbits, remedy the drawbacks
and improve upon the advantages of the SALI. For example, if we
could define an index that depends on {\it several} Lyapunov
exponents, this might accelerate considerably the identification
of chaotic motion.

\section{Definition of the GALI}
\label{GALI_d}

Let us consider an autonomous Hamiltonian system of $N$ degrees of
freedom having a Hamiltonian function
\begin{equation}
H(q_1,q_2, \ldots, q_N,p_1,p_2, \ldots, p_N)=h=\mbox{constant}
\label{eq:Ham}
\end{equation}
where $q_i$ and $p_i$, $i=1,2,\ldots,N$ are the generalized
coordinates and conjugate momenta respectively. An orbit of this
system is defined by a vector $\vec{x}(t)=(q_1(t),q_2(t), \ldots,
q_N(t),p_1(t),p_2(t), \ldots, p_N(t))$, with $x_i=q_i$,
$x_{i+N}=p_i$, $i=1,2,\ldots,N$. The time evolution of this orbit
is governed by Hamilton equations of motion
\begin{equation}
\frac{d \vec{x}}{dt}= \vec{\mathcal{V}}(\vec{x})= \left(
\frac{\partial H}{\partial \vec{p}}\, ,  - \frac{\partial
H}{\partial \vec{q}} \right), \label{eq:Hameq}
\end{equation}
while the time evolution of an initial deviation vector
$\vec{w}(0)=(dx_1(0),\ldots, dx_{2N}(0)) $ from the $\vec{x}(t)$
solution of (\ref{eq:Hameq}) obeys the variational equations
\begin{equation}
\frac{d \vec{w}}{dt} = \textbf{M}(\vec{x}(t)) \,\vec{w} \, ,
\label{eq:var}
\end{equation}
where $\textbf{M}= \partial \vec{\mathcal{V}} / \partial \vec{x}$
is the Jacobian matrix of $\vec{\mathcal{V}}$.

The SALI is a quantity suitable for checking whether or not two
normalized deviation vectors $\hat{w}_1$, $\hat{w}_2$ (having norm
1), eventually become linearly dependent, by falling in the same
direction. The linear dependence of the two vectors is equivalent
to the vanishing  of the `area' of the parallelogram having as
edges the two vectors. Generalizing this idea we now follow the
evolution of $k$ deviation vectors $\hat{w}_1$, $\hat{w}_2$,
$\ldots$, $\hat{w}_k$, with $2\leq k \leq 2N$, and determine
whether these eventually become linearly dependent, by checking if
the `volume' of the parallelepiped having these vectors as edges
goes to zero. This volume will be computed as the norm of the
wedge product of these vectors (see Appendix \ref{Wedge} for a definition
of the wedge product).

All normalized deviation vectors $\hat{w}_i$, $i=1,2,\ldots,k$,
belong to the $2N$--dimensional tangent space of the Hamiltonian
flow. Using as a basis of this space the usual set of orthonormal
vectors
\begin{equation}
\hat{e}_1= (1,0,0,\ldots,0), \hat{e}_2= (0,1,0,\ldots,0), \ldots,
\hat{e}_{2N}= (0,0,0,\ldots,1) \label{eq:basis}
\end{equation}
any deviation vector $\hat{w}_i$ can be written as
\begin{equation}
\hat{w}_i=\sum_{j=1}^{2N} w_{ij} \hat{e}_j \,\,\, , \,\,\,
i=1,2,\ldots,k \label{eq:vec}
\end{equation}
where $w_{ij}$ are real numbers satisfying
\begin{equation}
\sum_{j=1}^{2N} w_{ij}^2 =1. \label{eq:1}
\end{equation}
Thus, equation (\ref{eq:w_matrix}) of Appendix \ref{Wedge} gives
\begin{equation}
\left[\begin{array}{c}
 \hat{w}_1 \\ \hat{w}_2 \\ \vdots \\
\hat{w}_k \end{array} \right] = \left[
\begin{array}{cccc}
w_{11} & w_{12} & \cdots & w_{1\, 2N} \\
w_{21} & w_{22} & \cdots & w_{2\, 2N} \\
\vdots & \vdots &  & \vdots \\
w_{k1} & w_{k2} & \cdots & w_{k\, 2N} \end{array} \right] \cdot
\left[\begin{array}{c}
 \hat{e}_1 \\ \hat{e}_2 \\ \vdots \\
\hat{e}_{2N} \end{array} \right].\,\,\,   \label{eq:matrix_e}
\end{equation}
Using then equation (\ref{eq:w_wedge}) the wedge product of these
$k$ deviation vectors takes the form
\begin{equation}
\hat{w}_1\wedge \hat{w}_2\wedge \cdots \wedge\hat{w}_k = \sum_{1
\leq i_1 <i_2 < \cdots < i_k \leq 2N} \left|
\begin{array}{cccc}
w_{1 i_1} & w_{1 i_2} & \cdots & w_{1 i_k} \\
w_{2 i_1} & w_{2 i_2} & \cdots & w_{2 i_k} \\
\vdots & \vdots &  & \vdots \\
w_{k i_1} & w_{k i_2} & \cdots & w_{k i_k} \end{array} \right|
\hat{e}_{i_1}\wedge \hat{e}_{i_2} \wedge \cdots \wedge
\hat{e}_{i_k} \,\,\, ,\label{eq:volume_2}
\end{equation}
where the sum is performed over all possible combinations of $k$
indices out of $2N$.

If at least two of the normalized deviation vectors $\hat{w}_i$,
$i=1,2,\ldots,k$ are linearly dependent, all the $k \times k$
determinants appearing in equation (\ref{eq:volume_2}) will become
zero making the `volume' vanish. Equivalently the quantity
\begin{equation}
\|\hat{w}_1\wedge \hat{w}_2\wedge \cdots \wedge\hat{w}_k \|=
\left\{\sum_{1 \leq i_1 < i_2 < \cdots < i_k \leq 2N} \left|
\begin{array}{cccc}
w_{1 i_1} & w_{1 i_2} & \cdots & w_{1 i_k} \\
w_{2 i_1} & w_{2 i_2} & \cdots & w_{2 i_k} \\
\vdots & \vdots &  & \vdots \\
w_{k i_1} & w_{k i_2} & \cdots & w_{k i_k} \end{array} \right|^2
\right\}^{1/2} \label{eq:norm}
\end{equation}
which we shall call the `norm' of the wedge product, will also
become zero. Thus, we define this important quantity as the
Generalized Alignment Index (GALI) of order $k$
\begin{equation}
\mbox{GALI}_k(t)=\| \hat{w}_1(t)\wedge \hat{w}_2(t)\wedge \cdots
\wedge\hat{w}_k(t) \|\,\, . \label{eq:GALI}
\end{equation}

In order to compute GALI$_k$, therefore, we need to follow the
evolution of an orbit with initial conditions $\vec{x}(0)$, using
equation (\ref{eq:Hameq}), as well as the evolution of $k$
initially linearly independent unit deviation vectors $\hat{w}_i$,
$i=1,2,\ldots,k$ using the variational equations (\ref{eq:var}).
At every time step, we normalize these deviation vectors to unity
and compute GALI$_k$ as the norm of their wedge product using
equation (\ref{eq:norm}).

Consequently, if GALI$_k(t)$ tends to zero, this would imply that
the volume of the parallelepiped having the vectors $\hat{w}_i$ as
edges also shrinks to zero, as at least one of the deviation
vectors becomes linearly dependent on the remaining ones. On the
other hand, if GALI$_k(t)$ remains far from zero, as t grows
arbitrarily, this would indicate the linear independence of the
deviation vectors and the existence of a corresponding
parallelepiped, whose volume is different from zero for all time.

\section{Theoretical results}
\label{GALI_b}

\subsection{Exponential decay of GALI for chaotic orbits}
\label{GALI_chaos}

In order to investigate the dynamics in the vicinity of a chaotic
orbit of the Hamiltonian system (\ref{eq:Ham}) with $N$ degrees of
freedom, let us first recall some known properties of the Lyapunov
characteristic exponents, following e.\ g.\ \cite{LL92,CAMTV05}.
It has been shown that the mean exponential rate of divergence
$\sigma\left(\vec{x}(0),\vec{w}\right)$ from a reference orbit
with initial condition $\vec{x}(0)$ given by
\begin{equation}
\sigma\left(\vec{x}(0),\vec{w}\right)=\lim_{t\rightarrow \infty}
\frac{1}{t}\, \ln  \frac{\|\vec{w}(t)\|}{\|\vec{w}(0)\|}\,\, , \label{eq:s_lim}
\end{equation}
exists and  is finite. Furthermore there is a $2N$--dimensional
basis $\{ \hat{u}_1, \hat{u}_2,\ldots, \hat{u}_{2N} \}$ of the
tangent space of the Hamiltonian flow so that
$\sigma\left(\vec{x}(0),\vec{w}\right)$ takes one of the $2N$
(possibly nondistinct) values
\begin{equation}
\sigma_i\left(\vec{x}(0)\right)=\sigma\left(\vec{x}(0),\hat{u}_i\right)\,\,\, ,
\,\,\, i=1,2,\ldots,2N
\label{eq:s_i}
\end{equation}
which are the Lyapunov characteristic exponents, ordered in size
as follows:
\begin{equation}
\sigma_1 \geq \sigma_2 \geq \ldots \geq \sigma_{2N}\,\, .
\label{eq:Lyap_size}
\end{equation}

These properties can be easily understood if the reference orbit
is an unstable periodic solution of period $T$. In this case, the
matrix $\textbf{M}$ of the variational equations (\ref{eq:var}) is
a continuous $T$--periodic $2N\times 2N$ matrix. The solution of
equations (\ref{eq:var}) can be written as
\begin{equation}
\vec{w}(t)= \mbox{\boldmath $\Phi$}  (t)\cdot \vec{w}(0)\,\, , \label{eq:F}
\end{equation}
where $\mbox{\boldmath $\Phi$}(t)$ is the so--called fundamental
matrix (see e.\ g.\ \cite{V90}), such that $\mbox{\boldmath
$\Phi$}(0)=\textbf{I}$, the $2N\times 2N$ identity matrix. The
behavior of the deviation vector $\vec{w}(t)$ and consequently the
stability of the periodic orbit is determined by the eigenvalues
$\lambda_i$ of the so--called monodromy matrix $\mbox{\boldmath
$\Phi$}(T)$, ordered as $|\lambda_1| \geq |\lambda_2| \geq \cdots
\geq |\lambda_{2N}|$. Let $\hat{u}_i$, $i=1,2,\ldots, 2N$ denote
the corresponding eigenvectors. Then for $\vec{w}(0)=\hat{u}_i$ we
have
\begin{equation}
\vec{w}(nT)= \lambda_i^n  \hat{u}_i\,\,\, , \,\,\, i=1,2,\ldots
,2N\label{eq:w_lambda}
\end{equation}
and from (\ref{eq:s_lim}) we get
\begin{equation}
\sigma\left(\vec{x}(0),\hat{u}_i\right) = \lim_{t\rightarrow \infty}
\frac{1}{nT} \ln |\lambda_i^n|=\frac{\ln |\lambda_i|}{T}\,\,\, ,
\,\,\, i=1,2,\ldots,2N. \label{eq:sigma_lambda}
\end{equation}
Furthermore, if we write
\begin{equation}
\vec{w} (0) = \sum_{i=1}^{2N} c_i\,  \hat{u}_i
\, ,\label{eq:w0}
\end{equation}
it follows from (\ref{eq:w_lambda}) that the first nonvanishing
coefficient $c_i$ dominates the subsequent evolution of $\vec{w}(nT)$.
Thus, if $c_1 \neq 0$ we get from (\ref{eq:s_lim})
$\sigma\left(\vec{x}(0),\vec{w}\right)=\sigma_1$, if $c_1=0$ and $c_2 \neq 0$
we get $\sigma\left(\vec{x}(0),\vec{w}\right)=\sigma_2$ and so on.
So, the evolution of the initial deviation vector $\vec{w}(0)$ is well
approximated by
\begin{equation}
\vec{w} (nT) = \sum_{i=1}^{2N} c_i\, e^{\sigma_i n T} \hat{u}_i
\, ,\label{eq:wnT_evol}
\end{equation}

For a nonperiodic orbit we cannot define such eigenvalues and
eigenvectors as above. Nevertheless, Oseledec \cite{O68} has proven
the existence of  basis vectors $\{ \hat{u}_1, \hat{u}_2,\ldots,
\hat{u}_{2N} \}$ and  Lyapunov exponents for nonperiodic orbits.
This is perhaps not surprising, since periodic orbits are dense in
the phase space of Hamiltonian systems and thus a periodic orbit of
arbitrary large period can always be found arbitrary close to any
nonperiodic orbit. So, the time evolution of a deviation vector may
be approximated by a variant of equation (\ref{eq:wnT_evol}), i.~e.
\begin{equation}
\vec{w} (t) = \sum_{i=1}^{2N} c_i\, e^{d_i t} \hat{u}_i
\, ,\label{eq:wnT_variant}
\end{equation}
where $c_i$, $d_i$ are real numbers depending on the specific
phase space location through which the reference orbit passes.
Thus, the quantities $d_i$, $i=1,2,\ldots,2N$ may be thought of as
`local Lyapunov exponents' having as limits for $t \rightarrow
\infty$ the `global' LCEs $\sigma_i$, $i=1,2,\ldots,2N$. We notice
that even if in some special cases where the vectors $\hat{u}_i$,
$i=1,2,\ldots,2N$ are known a priori, so that one could set
$\vec{w}(0)= \hat{u}_i$, the computational errors in the numerical
evolution of the deviation vector would lead to the actual
computation of $\sigma_1$ from equation (\ref{eq:lyap1_def})
\cite{BGGS80b}.

It is well known that Hamiltonian systems are generically
non--integrable and possess Lyapunov exponents in chaotic domains
which are real and grouped in pairs of opposite sign with two of
them being equal to zero. We, therefore, have
$\sigma_i=-\sigma_{2N-i+1}$ for $i=1,2,\ldots,N$ and $\sigma_1
\geq \sigma_2 \geq \cdots \geq \sigma_{N-1} \geq
\sigma_{N}=\sigma_{N+1}=0 \geq \sigma_{N+2} \geq \cdots \geq
\sigma_{2N}$. Assuming that, after a certain time interval, the
$d_i$, $i=1,2,\ldots,2N$  do not fluctuate significantly about
their limiting values, we  write $d_i \approx \sigma_i$ and
express the evolution of the deviation vectors $\vec{w}_i$ in the
form
\begin{equation}
\vec{w}_i (t) = \sum_{j=1}^{2N} c_j^i\, e^{\sigma_j t}\, \hat{u}_j
\, \label{eq:dev_1}
\end{equation}
(see discussion in section \ref{App_2D} and figure
\ref{fig:2D_ch_L1}). Thus, if $\sigma_1 > \sigma_2$, a leading
order estimate of the deviation vector's Euclidean norm (for $t$
large enough), is given by
\begin{equation}
\|\vec{w}_i(t)\| \approx |c_1^i| e^{\sigma_1 t} \, .
\label{eq:vector_norm}
\end{equation}
Consequently, the matrix $\textbf{C}$ in (\ref{eq:w_matrix}) of
coefficients of $k$ normalized deviation vectors $\hat{w}_i(t)=
\vec{w}_i(t)/\|\vec{w}_i(t)\|$, $i=1,2,\ldots,k$ with $2\leq k\leq
2N$, using as basis of the vector space the set $\{ \hat{u}_1,
\hat{u}_2,\ldots, \hat{u}_{2N} \}$ becomes
\begin{equation}
\textbf{C}(t)=\left[c_{ij}\right]= \left[ \begin{array}{ccccc} s_1
& \frac{c_2^1}{|c_1^1|} e^{-(\sigma_1-\sigma_2)t} &
\frac{c_3^1}{|c_1^1|} e^{-(\sigma_1-\sigma_3)t} & \cdots &
\frac{c_{2N}^1}{|c_1^1|} e^{-(\sigma_1-\sigma_{2N})t} \\
s_2 & \frac{c_2^2}{|c_1^2|} e^{-(\sigma_1-\sigma_2)t} &
\frac{c_3^2}{|c_1^2|} e^{-(\sigma_1-\sigma_3)t} & \cdots &
\frac{c_{2N}^2}{|c_1^2|} e^{-(\sigma_1-\sigma_{2N})t} \\
\vdots & \vdots & \vdots & & \vdots \\
s_k & \frac{c_2^k}{|c_1^k|} e^{-(\sigma_1-\sigma_2)t} &
\frac{c_3^k}{|c_1^k|} e^{-(\sigma_1-\sigma_3)t} & \cdots &
\frac{c_{2N}^k}{|c_1^k|} e^{-(\sigma_1-\sigma_{2N})t}
\end{array} \right] ,
\label{eq:matix _C}
\end{equation}
with $s_i=\mbox{sign}(c_1^i)$  and $i=1,2,\ldots,k$,
$j=1,2,\ldots, 2N$ and so we have
\begin{equation}
\left[ \begin{array}{cccc} \hat{w}_1 &\hat{w}_2 &\ldots
&\hat{w}_{k} \end{array}\right]^{\mathrm{T}} =
\textbf{C}\cdot\left[
\begin{array}{cccc} \hat{u}_1 &\hat{u}_2 &\ldots &\hat{u}_{2N}
\end{array}\right]^{\mathrm{T}} \label{eq:w_u}
\end{equation}
with ($^{\mathrm{T}}$) denoting the transpose of a matrix. The
wedge product of the $k$ normalized deviation vectors is then
computed as in equation (\ref{eq:volume_2}) by:
\begin{equation}
\hat{w}_1(t)\wedge \hat{w}_2(t)\wedge \cdots \wedge\hat{w}_k(t) =
\sum_{1 \leq i_1 < i_2 < \cdots < i_k \leq 2N} \left|
\begin{array}{cccc}
c_{1 i_1} & c_{1 i_2} & \cdots & c_{1 i_k} \\
c_{2 i_1} & c_{2 i_2} & \cdots & c_{2 i_k} \\
\vdots & \vdots &  & \vdots \\
c_{k i_1} & c_{k i_2} & \cdots & c_{k i_k} \end{array} \right|
\hat{u}_{i_1}\wedge \hat{u}_{i_2} \wedge \cdots \wedge
\hat{u}_{i_k}. \label{eq:ch_wedge}
\end{equation}

Note that the quantity
\begin{equation}
S_k= \left\{\sum_{1 \leq i_1 < i_2 < \cdots < i_k \leq 2N} \left|
\begin{array}{cccc}
c_{1 i_1} & c_{1 i_2} & \cdots & c_{1 i_k} \\
c_{2 i_1} & c_{2 i_2} & \cdots & c_{2 i_k} \\
\vdots & \vdots &  & \vdots \\
c_{k i_1} & c_{k i_2} & \cdots & c_{k i_k} \end{array}
\right|^2\right\}^{1/2} \label{eq:ch_sk}
\end{equation}
is {\it not identical} to the norm (\ref{eq:norm}) of the
$k$--vector $\hat{w}_1(t)\wedge \hat{w}_2(t)\wedge \cdots
\wedge\hat{w}_k(t)$ as the wedge product in equation
(\ref{eq:ch_wedge}) is not expressed with respect to the basis
(\ref{eq:basis}). Thus one should consider the transformation
\begin{equation}
\left[ \begin{array}{cccc} \hat{u}_1 &\hat{u}_2 &\ldots
&\hat{u}_{2N} \end{array}\right]^{\mathrm{T}} =
\textbf{T}_c\cdot\left[
\begin{array}{cccc} \hat{e}_1 &\hat{e}_2 &\ldots &\hat{e}_{2N}
\end{array}\right]^{\mathrm{T}}, \label{eq:transform_basis}
\end{equation}
between the two bases, with $\textbf{T}_c$ denoting the
transformation matrix. Of course, when considering the wedge
product of $2N$ deviation vectors one can easily show that
\begin{equation} \label{eq:2N_basis}
\|\hat{w}_1\wedge \hat{w}_2\wedge \cdots
\wedge\hat{w}_{2N}\|=S_{2N}\cdot |\mbox{det}\textbf{T}_c|\, .
\end{equation}

If, on the other hand, we consider the wedge product of {\it
fewer} than $2N$ deviation vectors, the norm (\ref{eq:norm}) and
the quantity $S_k$ (\ref{eq:ch_sk}) are not related through a
simple expression like (\ref{eq:2N_basis}). We shall proceed,
however, to obtain results using (\ref{eq:ch_sk}) instead of
(\ref{eq:norm}), as we do not expect that such a change of basis
will affect significantly the dynamics and alter our conclusions
for the following reasons: First, we note that both quantities are
zero when at least two of the $k$ deviation vectors are linearly
dependent, due to the fact that all the determinants appearing in
equations (\ref{eq:norm}) and (\ref{eq:ch_sk}) vanish. In
addition, the transformation matrix $\textbf{T}_c$ is not singular
as the sets $\{\hat{u}_i\}$ and $\{\hat{e}_i\}$, $i=1,2,\ldots,2N$
continue to be valid bases of the vector space. Thus, both
quantities are expected to behave in a similar way in the case of
chaotic orbits, where  the deviation vectors tend to become
linearly dependent. Thus, by studying analytically the time
evolution of $S_k$ through (\ref{eq:ch_sk}), we expect to derive
accurate approximations of the behavior of the GALI$_k$
(\ref{eq:GALI}) for chaotic orbits. The validity of this
approximation is numerically tested and verified in section
\ref{GALI_a}.

Let us now see how this approximation is derived: The determinants
appearing in the definition of $S_k$ (see equation (\ref{eq:ch_sk}))
can be divided in two categories depending on whether or not they
contain the first column of matrix $\textbf{C}$. Using standard
properties of determinants, we see that those that do contain the
first column yield
\begin{eqnarray}
\label{eq:dets_1}  D_{1,j_1,j_2,\ldots,j_{k-1}}= \left|
\begin{array}{cccc}
s_1 & \frac{c_{j_1}^1}{|c_1^1|} e^{-(\sigma_1-\sigma_{j_1})t} &
 \cdots &
\frac{c_{j_{k-1}}^1}{|c_1^1|} e^{-(\sigma_1-\sigma_{j_{k-1}})t} \\
s_2 & \frac{c_{j_1}^2}{|c_1^2|} e^{-(\sigma_1-\sigma_{j_1})t} &
 \cdots &
\frac{c_{j_{k-1}}^2}{|c_1^2|} e^{-(\sigma_1-\sigma_{j_{k-1}})t} \\
\vdots & \vdots  & & \vdots \\
s_k & \frac{c_{j_1}^k}{|c_1^k|} e^{-(\sigma_1-\sigma_{j_1})t} &
 \cdots &
\frac{c_{j_{k-1}}^k}{|c_1^k|} e^{-(\sigma_1-\sigma_{j_{k-1}})t}
\end{array} \right| =
\nonumber\\
= \left|
\begin{array}{cccc}
s_1 & \frac{c_{j_1}^1}{|c_1^1|}  &
 \cdots &
\frac{c_{j_{k-1}}^1}{|c_1^1|}  \\
s_2 & \frac{c_{j_1}^2}{|c_1^2|}  &
 \cdots &
\frac{c_{j_{k-1}}^2}{|c_1^2|}  \\
\vdots & \vdots  & & \vdots \\
s_k & \frac{c_{j_1}^k}{|c_1^k|}  &
 \cdots &
\frac{c_{j_{k-1}}^k}{|c_1^k|} \cdot
\end{array} \right|e^{-\left[
(\sigma_1-\sigma_{j_1}) + (\sigma_1-\sigma_{j_2})+ \cdots+
(\sigma_1-\sigma_{j_{k-1}})\right] t}
\end{eqnarray}
with $1<j_1<j_2<\ldots < j_{k-1} \leq 2N$. Thus, the time
evolution of $D_{1,j_1,j_2,\ldots,j_{k-1}}$ is mainly determined
by the exponential law
\begin{equation}\label{eq:time_ev_det_1}
D_{1,j_1,j_2,\ldots,j_{k-1}} \propto e^{-\left[
(\sigma_1-\sigma_{j_1}) + (\sigma_1-\sigma_{j_2})+ \cdots+
(\sigma_1-\sigma_{j_{k-1}})\right] t}\, .
\end{equation}

Similarly, we deduce that the determinants that {\it do not}
contain the first column of matrix $\textbf{C}$ (\ref{eq:matix
_C}) have the form
\begin{eqnarray}
\label{eq:dets_2}  D_{j_1,j_2,\ldots,j_{k}}= \left|
\begin{array}{cccc}
\frac{c_{j_1}^1}{|c_1^1|} e^{-(\sigma_1-\sigma_{j_1})t} &
\frac{c_{j_2}^1}{|c_1^1|} e^{-(\sigma_1-\sigma_{j_2})t} &
 \cdots &
\frac{c_{j_{k}}^1}{|c_1^1|} e^{-(\sigma_1-\sigma_{j_{k}})t}\\
\frac{c_{j_1}^2}{|c_1^2|} e^{-(\sigma_1-\sigma_{j_1})t} &
\frac{c_{j_2}^2}{|c_1^2|} e^{-(\sigma_1-\sigma_{j_2})t} &
 \cdots &
\frac{c_{j_{k}}^2}{|c_1^2|} e^{-(\sigma_1-\sigma_{j_{k}})t} \\
\vdots & \vdots  & & \vdots \\
\frac{c_{j_1}^k}{|c_1^k|} e^{-(\sigma_1-\sigma_{j_1})t} &
\frac{c_{j_2}^k}{|c_1^k|} e^{-(\sigma_1-\sigma_{j_2})t} &
 \cdots &
\frac{c_{j_{k}}^k}{|c_1^k|} e^{-(\sigma_1-\sigma_{j_{k}})t}
\end{array} \right|=\nonumber\\
= \left|
\begin{array}{cccc}
\frac{c_{j_1}^1}{|c_1^1|}  & \frac{c_{j_2}^1}{|c_1^1|}  &
 \cdots &
\frac{c_{j_{k}}^1}{|c_1^1|}  \\
\frac{c_{j_1}^2}{|c_1^2|} & \frac{c_{j_2}^2}{|c_1^2|}  &
 \cdots &
\frac{c_{j_{k}}^2}{|c_1^2|}  \\
\vdots & \vdots  & & \vdots \\
\frac{c_{j_1}^k}{|c_1^k|}& \frac{c_{j_1}^k}{|c_1^k|}  &
 \cdots &
\frac{c_{j_{k-1}}^k}{|c_1^k|} \cdot
\end{array} \right|e^{-\left[
(\sigma_1-\sigma_{j_1}) + (\sigma_1-\sigma_{j_2})+
\cdots+(\sigma_1-\sigma_{j_{k-1}})+
(\sigma_1-\sigma_{j_{k}})\right] t}
\end{eqnarray}
with $1<j_1<j_2<\ldots < j_{k-1} <j_k \leq 2N$. Thus, the values
of these determinants also tend  to zero following an exponential
law
\begin{equation}\label{eq:time_ev_det_2}
D_{j_1,j_2,\ldots,j_{k}}  \propto  e^{-\left[
(\sigma_1-\sigma_{j_1}) + (\sigma_1-\sigma_{j_2})+
\cdots+(\sigma_1-\sigma_{j_{k-1}})+
(\sigma_1-\sigma_{j_{k}})\right] t}.
\end{equation}

Clearly, from all determinants appearing in the definition of
$S_k$, (\ref{eq:ch_sk}), the one that decreases the
\textit{slowest} is the one containing the first $k$ columns of
matrix $\textbf{C}$ in (\ref{eq:matix _C}):
\begin{equation}\label{eq:time_ev_det_123k}
D_{1,2,3,\ldots,k}  \propto e^{-\left[ (\sigma_1-\sigma_2) +
(\sigma_1-\sigma_3)+ \cdots+ (\sigma_1-\sigma_k)\right] t}\, .
\end{equation}
All other determinants appearing in equations
(\ref{eq:time_ev_det_1}) and (\ref{eq:time_ev_det_2}) tend to zero
\textit{faster} than $D_{1,2,3,\ldots,k}$ since the quantities in
their exponentials are smaller or equal to the exponent in
(\ref{eq:time_ev_det_123k}). We, therefore, conclude that the rate
of decrease of $S_k$  is dominated by (\ref{eq:time_ev_det_123k}),
yielding the approximation
\begin{equation}\label{eq:sk_decay}
S_k (t) \propto e^{-\left[ (\sigma_1-\sigma_2) +
(\sigma_1-\sigma_3)+ \cdots+ (\sigma_1-\sigma_k)\right] t}\, .
\end{equation}
Furthermore, since the norm (\ref{eq:norm}) of the $k$--vector
$\hat{w}_1 \wedge \hat{w}_2 \wedge \cdots \wedge \hat{w}_k$ is
expected to evolve in a similar way as $S_k$, we conclude that
GALI$_k$ tends to zero in the same manner as above, i.e.
\begin{equation}\label{eq:ch_GALI_decay}
\mbox{GALI}_k(t)  \propto e^{-\left[ (\sigma_1-\sigma_2) +
(\sigma_1-\sigma_3)+ \cdots+ (\sigma_1-\sigma_k)\right] t}\, .
\end{equation}
We note here that in \cite{SABV04}, where it was shown theoretically
that SALI tends exponentially to zero for chaotic orbits as
$\mbox{SALI}(t) \propto \exp\{-(\sigma_1-\sigma_2)t\}$ (which is
equivalent to equation (\ref{eq:ch_GALI_decay}) for $k=2$), equation
(\ref{eq:dev_1}) was also retrieved, although it was wrongly assumed
that the LCEs are related to the eigenvalues of matrix $\textbf{M}$
of the variational equations (\ref{eq:var}).

In the previous analysis we assumed that $\sigma_1 >
\sigma_2$ so that the norm of each deviation vector can be well
approximated by equation (\ref{eq:vector_norm}). If the first $m$
Lyapunov exponents, with $1<m<k$, are equal, or very close to each
other, i.e.~$\sigma_1 \simeq\sigma_2 \simeq \cdots \simeq
\sigma_m$ equation (\ref{eq:ch_GALI_decay}) becomes
\begin{equation}\label{eq:ch_GALI_decay_2}
\mbox{GALI}_k (t) \propto e^{-\left[ (\sigma_1-\sigma_{m+1}) +
(\sigma_1-\sigma_{m+2})+ \cdots+ (\sigma_1-\sigma_k)\right] t}\, ,
\end{equation}
which still describes an exponential decay. However, for $ k \leq
m < N$ the GALI$_k$ does not tend to zero as there exists at least
one determinant of the matrix $\textbf{C}$ that does not vanish.
In this case, of course, one should increase the number of
deviation vectors until an exponential decrease of GALI$_k$ is
achieved. The extreme situation that all $\sigma_i=0$ corresponds
to motion on quasiperiodic tori, where all orbits are regular and
is described below.

\subsection{The evaluation of GALI for regular orbits}
\label{GALI_order}

As is well--known, regular orbits of an $N$ degree of freedom
Hamiltonian system (\ref{eq:Ham}) typically lie on
$N$--dimensional tori. If such tori are found around a stable
periodic orbit, they can be accurately described by $N$ formal
integrals of motion in involution, so that the system would appear
locally integrable. This means that we could perform a local
transformation to action--angle variables, considering as actions
$J_1, J_2, \ldots, J_N$ the values of the $N$ formal integrals, so
that Hamilton's equations of motion,  locally attain the form
\begin{equation}
\begin{array}{ccl}
\dot{J}_i & = & 0 \\
\dot{\theta}_i & = & \omega_i(J_1, J_2, \ldots, J_N)
\end{array}
\,\,\, i=1,2,\ldots,N. \label{eq:a-a}
\end{equation}
These can be easily integrated to give
\begin{equation}
\begin{array}{ccl}
J_i(t) & = & J_{i0} \\
\theta_i(t) & = &\theta_{i0} +\omega_i(J_{10}, J_{20}, \ldots,
J_{N0})\, t
\end{array}
\,\,\, i=1,2,\ldots,N, \label{eq:a-a-sol}
\end{equation}
where $J_{i0}$, $\theta_{i0}$, $i=1,2,\ldots,N$ are the initial
conditions.

By denoting as $\xi_i$, $\eta_i$, $i=1,2,\ldots,N$  small
deviations of $J_i$ and $\theta_i$ respectively, the variational
equations of system (\ref{eq:a-a}), describing the evolution of a
deviation vector are
\begin{equation}
\begin{array}{ccl}
\dot{\xi}_i & = & 0 \\
\dot{\eta}_i & = & \sum_{j=1}^N \omega_{ij}\cdot \xi_j
\end{array}\,\,\,
i=1,2,\ldots,N, \label{eq:twist_map_var}
\end{equation}
where
\begin{equation}
\omega_{ij}=\frac{\partial \omega_i}{\partial
J_j}\left|_{\vec{J}_0} \right. \,\,\,
i,j=1,2,\ldots,N,\label{eq:omega}
\end{equation}
and $\vec{J}_0=(J_{10}, J_{20}, \ldots, J_{N0})=\mbox{constant}$,
represents the $N$--dimensional vector of the initial actions. The
solution of these equations is:
\begin{equation}
\begin{array}{ccl}
\xi_i(t) & = & \xi_i(0) \\
\eta_i (t) & = & \eta_i (0) + \left[ \sum_{j=1}^N \omega_{ij}
\xi_j(0)\right] \, t
\end{array} \,\,\, i=1,2,\ldots,N.
\label{eq:twist_map_var_sol}
\end{equation}

From equations (\ref{eq:twist_map_var_sol}) we see that an initial
deviation vector $\vec{w}(0)$ with coordinates $\xi_i(0)$,
$i=1,2,\ldots,N$ in the action variables and $\eta_i(0)$,
$i=1,2,\ldots,N$ in the angles,
i.~e.~$\vec{w}(0)=(\xi_1(0),\xi_2(0),\ldots,
\xi_N(0),\eta_1(0),\eta_2(0),\ldots, \eta_N(0))$, evolves in time
in such a way that its action coordinates remain constant, while
its angle coordinates increase linearly in time. This behavior
implies an almost linear increase of the norm of the deviation
vector. To see this, let us assume that initially this vector
$\vec{w}(0)$ has unit magnitude, i.~e.
\begin{equation}
\sum_{i=1}^N\xi_i(0)^2 +  \sum_{i=1}^N\eta_i(0)^2 =1
\label{eq:twist_map_unit}
\end{equation}
whence the time evolution of its norm is given by
\begin{equation}
\|\vec{w}(t)\|= \left\{ 1+ \left[\sum_{i=1}^N \left( \sum_{j=1}^N
\omega_{ij} \xi_j(0)  \right)^2 \right] t^2 + \left[2\sum_{i=1}^N
\left(\eta_i(0) \sum_{j=1}^N  \omega_{ij} \xi_j(0)  \right)
 \right] t\right\} ^{1/2}, \label{eq:twist_map_norm}
\end{equation}
while the normalized deviation vector $\hat{w}(t)$ becomes:
\begin{equation}
\hat{w}(t) = \frac{1}{\|\vec{w}(t)\|} \left(\xi_1(0), \ldots,
\xi_N(0),\eta_1(0)+\left[ \sum_{j=1}^N \omega_{1j} \xi_j(0)\right]
t,\ldots, \eta_N(0)+\left[ \sum_{j=1}^N \omega_{Nj}
\xi_j(0)\right] t \right). \label{eq:twist_map_norm_vec}
\end{equation}
Since the norm (\ref{eq:twist_map_norm}) of a deviation vector,
for $t$ large enough, increases practically linearly with t, the
normalized deviation vector (\ref{eq:twist_map_norm_vec}) tends to
fall on the tangent space of the torus, since its coordinates
perpendicular to the torus (i.~e.~the coordinates along the action
directions) vanish following a $t^{-1}$ rate. This behavior has
already been shown numerically in the case of an integrable
Hamiltonian of 2 degrees of freedom in \cite{SABV03b}.

Using as a basis of the $2N$--dimensional tangent space of the
Hamiltonian flow the $2N$ unit vectors
$\{\hat{v}_1,\hat{v}_2,\ldots,\hat{v}_{2N}\}$, such that the first
$N$ of them, $\hat{v}_1,\hat{v}_2,\ldots,\hat{v}_{N}$ correspond
to the $N$ action variables and the remaining ones,
$\hat{v}_{N+1},\hat{v}_{N+2},\ldots,\hat{v}_{2N}$ to the $N$
conjugate angle variables, any unit deviation vector $\hat{w}_i$,
$i=1,2,\ldots$ can be written as
\begin{equation}
\hat{w}_i(t) = \frac{1}{\|\vec{w}(t)\|}  \left[\sum_{j=1}^{N}
\xi_j^i(0) \, \hat{v}_j + \sum_{j=1}^{N}  \left(\eta_j^i(0)+
\sum_{k=1}^N \omega_{kj} \xi_j^i(0) t\right)   \hat{v}_{N+j}
\right]. \label{eq:order_dev_vec}
\end{equation}
We point out that the quantities $\omega_{ij}$, $i,j=1,2\ldots,N$,
in (\ref{eq:omega}), depend only on the particular reference orbit
and not on the choice of the deviation vector. We also note that
the basis $\hat{u}_i$, $i=1,2,\ldots, 2N$ depends on the specific
torus on which the motion occurs and is related to the usual
vector basis $\hat{e}_i$, $i=1,2,\ldots, 2N$ of equation
(\ref{eq:basis}), through a non--singular transformation, similar
to the one of equation (\ref{eq:transform_basis}), having the
form:
\begin{equation}
\left[ \begin{array}{cccc} \hat{v}_1 &\hat{v}_2 &\ldots
&\hat{v}_{2N} \end{array}\right]^{\mathrm{T}} =
\textbf{T}_o\cdot\left[
\begin{array}{cccc} \hat{e}_1 &\hat{e}_2 &\ldots &\hat{e}_{2N}
\end{array}\right]^{\mathrm{T}} \label{eq:transform_basis_order}
\end{equation}
with $\textbf{T}_o$ denoting the transformation matrix. The basis
$\{\hat{e}_1,\hat{e}_2,\ldots,\hat{e}_{2N}\}$ is used to describe
the evolution of a deviation vector with respect to the original
$q_i$, $p_i$ $i=1,2,\ldots, N$ coordinates of the Hamiltonian
system (\ref{eq:Ham}), while the basis
$\{\hat{v}_1,\hat{v}_2,\ldots,\hat{v}_{2N}\}$ is used to describe
the same evolution, if we consider the original system  in
action--angle variables, so that the equations of motion are the
ones given by (\ref{eq:a-a}).

At this point we make the following remark: If the initial
deviation vector already {\it lies in the tangent space of the
torus} it will remain constant for all time! Indeed, taking for
the initial conditions of this vector
\begin{equation}
\xi_i(0)=0, \,\,\,  i=1,2,\ldots,N \,\,\, ,
\label{eq:tangent_ini_1}
\end{equation}
with
\begin{equation}
\sum_{i=1}^N\eta_i(0)^2 =1, \label{eq:tangent_ini_2}
\end{equation}
we conclude from equation (\ref{eq:twist_map_var_sol}) that
\begin{equation}
\xi_i(t)=0 \,\,\, , \,\,\, \eta_i(t) =\eta_i(0).
\label{eq:tangent_evol}
\end{equation}
i.e. the deviation vector remains unchanged having its norm always
equal to 1. In particular, such a vector has the form
\begin{equation}
\hat{w}(t) =  \left(0,0, \ldots, 0,\eta_1(0),
\eta_2(0),\ldots,\eta_N(0)\right). \label{eq:twist_map_norm_vec_2}
\end{equation}

Let us now study the case of $k$, general, linearly independent
unit deviation vectors
$\{\hat{w}_1,\hat{w}_2,\ldots,\hat{w}_{k}\}$ with $2\leq k \leq
2N$. Using as vector basis the set
$\{\hat{v}_1,\hat{v}_2,\ldots,\hat{v}_{2N}\}$ we get:
\begin{equation}
\left[ \begin{array}{cccc} \hat{w}_1 &\hat{w}_2 &\ldots
&\hat{w}_{k} \end{array}\right]^{\mathrm{T}} =
\textbf{D}\cdot\left[
\begin{array}{cccc} \hat{v}_1 &\hat{v}_2 &\ldots &\hat{v}_{2N}
\end{array}\right]^{\mathrm{T}} \label{eq:w_v}
\end{equation}
If {\it no} deviation vector is initially located in the tangent
space of the torus, matrix $\textbf{D}$ has the form
\begin{eqnarray}
 \textbf{D} = \left[d_{ij}\right]  = \frac{1}{\prod_{m=1}^k
\|\vec{w}_m(t)\|}\cdot & & \nonumber \\  \cdot
\left[\begin{array}{cccccc} \xi_1^1(0) &  \cdots & \xi_N^1(0) &
\eta_1^1(0)+\sum_{m=1}^N \omega_{1m}\xi_m^1(0) t & \cdots &
\eta_N^1(0)+\sum_{m=1}^N \omega_{Nm}\xi_m^1(0) t \\
\xi_1^2(0) &  \cdots & \xi_N^2(0) & \eta_1^2(0)+\sum_{m=1}^N
\omega_{1m}\xi_m^2(0) t & \cdots &
\eta_N^2(0)+\sum_{m=1}^N \omega_{Nm}\xi_m^2(0) t \\
\vdots &  & \vdots & \vdots & & \vdots\\
\xi_1^k(0) &  \cdots & \xi_N^k(0) & \eta_1^k(0)+\sum_{m=1}^N
\omega_{1m}\xi_m^k(0) t & \cdots & \eta_N^k(0)+\sum_{m=1}^N
\omega_{Nm}\xi_m^k(0) t  \end{array} \right], & &\,\,\,\,\,\,\,\,
\label{eq:matix _D1a}
\end{eqnarray}
where $i=1,2,\ldots,k$ and $j=1,2,\ldots,2N$. Recalling our
earlier discussion (see
(\ref{eq:twist_map_unit})-(\ref{eq:order_dev_vec})), we note that
the norm of vector $\vec{w}_i(t)$ for long times, grows linearly
with t as
\begin{equation}
M_i (t)= \|\vec{w}_i(t)\|\propto t .  \label{eq:Mi}
\end{equation}
Defining then by $\mbox{\boldmath $\xi$}_i^{0,k}$ and
$\mbox{\boldmath $\eta$}_i^{k}$ the $k\times 1$ column matrices
\begin{equation}
\mbox{\boldmath $\xi$}_i^{0,k} =
\left[ \begin{array}{cccc} \xi_i^1(0) &\xi_i^2(0) &\ldots
&\xi_i^k(0) \end{array}\right]^{\mathrm{T}}
 \,\,\, , \,\,\,
\mbox{\boldmath $\eta$}_i^{k} =
\left[ \begin{array}{cccc} \eta_i^1(0) &\eta_i^2(0) &\ldots
&\eta_i^k(0) \end{array}\right]^{\mathrm{T}}, \label{eq:xi_eta}
\end{equation}
the matrix $\textbf{D}$ of (\ref{eq:matix _D1a}) assumes the much
simpler form
\begin{eqnarray}
\textbf{D}(t)  =  \frac{1}{\prod_{i=1}^k M_i(t)} \cdot \left[
\begin{array}{cccccc}
\mbox{\boldmath $\xi$}_1^{0,k} & \ldots & \mbox{\boldmath
$\xi$}_N^{0,k} & \mbox{\boldmath$\eta$}_1^{k} + \sum_{i=1}^N
\omega_{1i} \mbox{\boldmath $\xi$}_i^{0,k}t &  \ldots &
\mbox{\boldmath $\eta$}_N^{k} + \sum_{i=1}^N
\omega_{Ni}\mbox{\boldmath $\xi$}_i^{0,k}t
\end{array} \right] =\nonumber & & \\
 = \frac{1}{\prod_{i=1}^k M_i(t)} \cdot \textbf{D}^{0,k}(t). & & \,\,\,\,\,\,\,\,\,\,\,
\label{eq:matix _D1b}
\end{eqnarray}
Suppose now that we have  $m$ linearly independent deviation
vectors, with $m \leq k$ and $m\leq N$, initially located in the
tangent space of the torus and let them be the first $m$ deviation
vectors in equation (\ref{eq:w_v}). This implies, in the above
notation, that the $\mbox{\boldmath $\xi$}_i$ vectors in
(\ref{eq:matix _D1b}) now have the form
\begin{equation}
\mbox{\boldmath $\xi$}_i^{m,k} =
\left[ \begin{array}{cccccccc} 0 & 0 & \ldots & 0 &\xi_i^{m+1}(0)
&\xi_i^{m+2}(0) &\ldots
&\xi_i^k(0) \end{array}\right]^{\mathrm{T}}
\label{eq:xi_0}
\end{equation}
where the first superscript, $m$, refers to the number of first
components being equal to zero. Thus, the matrix $\textbf{D}$ of
(\ref{eq:matix _D1b}) in this case reads
\begin{eqnarray}
\textbf{D}(t)  = \frac{1}{\prod_{i=1}^{k-m} M_{m+i}(t)}
\cdot \left[
\begin{array}{cccccc}
\mbox{\boldmath $\xi$}_1^{m,k} & \ldots & \mbox{\boldmath
$\xi$}_N^{m,k} & \mbox{\boldmath$\eta$}_1^{k} + \sum_{i=1}^N
\omega_{1i} \mbox{\boldmath $\xi$}_i^{m,k}t &  \ldots &
\mbox{\boldmath $\eta$}_N^{k} + \sum_{i=1}^N
\omega_{Ni}\mbox{\boldmath $\xi$}_i^{m,k}t
\end{array} \right] =\nonumber & & \\
 = \frac{1}{\prod_{i=1}^{k-m} M_{m+i}(t)} \cdot \textbf{D}^{m,k}(t), & &
 \,\,\,\,\,\,\,\,\,\,\,
\label{eq:matix _D2a}
\end{eqnarray}
where the first superscript of $\textbf{D}^{m,k}(t)$ in equations
(\ref{eq:matix _D1b}) and (\ref{eq:matix _D2a}) has an
analogous meaning as in the $\mbox{\boldmath $\xi$}_i^{m,k}$. We note
that for $k=m$ we define $\prod_{i=1}^{0} M_{m+i}(t)=1$.

Using again equation (\ref{eq:w_wedge}), we write the wedge
product of the $k$ normalized deviation vectors as
\begin{equation}
\hat{w}_1(t)\wedge \hat{w}_2(t)\wedge \cdots \wedge\hat{w}_k(t) =
\sum_{1 \leq i_1 < i_2 < \cdots < i_k \leq 2N} \left|
\begin{array}{cccc}
d_{1 i_1} & d_{1 i_2} & \cdots & d_{1 i_k} \\
d_{2 i_1} & d_{2 i_2} & \cdots & d_{2 i_k} \\
\vdots & \vdots &  & \vdots \\
d_{k i_1} & d_{k i_2} & \cdots & d_{k i_k} \end{array} \right|
\hat{u}_{i_1}\wedge \hat{u}_{i_2} \wedge \cdots \wedge
\hat{u}_{i_k}. \label{eq:ord_wedge}
\end{equation}
and introduce the analogous quantity
\begin{equation}
S_k'= \left\{\sum_{1 \leq i_1 < i_2 < \cdots < i_k \leq 2N} \left|
\begin{array}{cccc}
d_{1 i_1} & d_{1 i_2} & \cdots & d_{1 i_k} \\
d_{2 i_1} & d_{2 i_2} & \cdots & d_{2 i_k} \\
\vdots & \vdots &  & \vdots \\
d_{k i_1} & d_{k i_2} & \cdots & d_{k i_k} \end{array}
\right|^2\right\}^{1/2}. \label{eq:ord_sk}
\end{equation}
as in the case of chaotic orbits, see (\ref{eq:ch_wedge}) and
(\ref{eq:ch_sk}) respectively.

As we have already explained, the $k$ deviation vectors will
eventually fall on the $N$--dimensional tangent space of the torus
on which the motion occurs. Of course, if some of them are already
located in the tangent space, at $t=0$, they will remain there
forever.  In their final state, the deviation vectors will have
coordinates only in the $N$--dimensional space spanned by
$\hat{v}_{N+1},\hat{v}_{N+2},\ldots,\hat{v}_{2N}$. Now, if we
start with $2\leq k \leq N$ general deviation vectors there is no
particular reason for them to become linearly dependent and their
wedge product will be different from zero, yielding $S_k'$ and
GALI$_k$ which are {\it not zero}. However, if we start with $N< k
\leq 2N$ deviation vectors, some of them will necessarily become
linearly dependent. Thus, in this case, their wedge product (as
well as $S_k'$ and GALI$_k$) will be zero.

We, therefore, need to examine in more detail the behavior of
these $S_k'$. Since, in general, we choose the initial deviation
vectors randomly (insisting only that they be linearly
independent), the most common situation is that none of the
initial deviation vectors is tangent to the torus. However, as we
are not certain that this will always hold, let us suppose that
$0<m\leq N$ of our deviation vectors are initially in the tangent
space of the torus. For $2\leq k \leq N$, this will make no
difference, as the GALI$_k$ tends to a non-zero constant. However,
for $N<k\leq2N$, GALI$_k$ goes to zero by a power law and the fact
that $m$ vectors are already in the tangent space, at $t=0$, may
significantly affect the decay rate of the index. Thus, in such
cases, the behavior of GALI needs to be treated separately.

\subsubsection{The case of $m=0$ tangent initial deviation vectors}
\label{no tangent vectors}

Let us consider first the most general case that {\it no}
deviation vector is initially  tangent to the torus. In this case,
the matrix $\textbf{D}$, whose elements appear in the definition
of $S_k'$, has the form given in equation (\ref{eq:matix _D1b}).
So all determinants appearing in the definition of $S_k'$ have as
a common factor the quantity $1/ \prod_{i=1}^k M_i(t)$, which, due
to (\ref{eq:Mi}), decreases to zero according to the power law
\begin{equation}
\frac{1}{\prod_{i=1}^k M_i(t)} \propto \frac{1}{t^k}
.\label{eq:M_prop_tk}
\end{equation}
In order to determine the precise time evolution of $S_k'$, we
search for the fastest increasing determinants of all the possible
$k \times k$ minors of the matrix $\textbf{D}^{0,k}$, in
(\ref{eq:matix _D1b}), as time $t$ grows.

Let us start with $k$ being less than or equal to the dimension of
the tangent space of the torus, i.~e.~$2\leq k \leq N$. The
fastest increasing determinants in this case are the
$N!/(k!(N-k)!)$ determinants, whose $k$ columns are chosen among
the last $N$ columns of matrix $\textbf{D}^{0,k}$:
\begin{equation}
\Delta_{j_1,j_2,\ldots,j_{k}}^{0,k}=\left|
\begin{array}{cccc}
\mbox{\boldmath$\eta$}_{j_1}^{k} + \sum_{i=1}^N \omega_{j_{1}i}
\mbox{\boldmath $\xi$}_i^{0,k}t & \mbox{\boldmath$\eta$}_{j_2}^{k}
+ \sum_{i=1}^N \omega_{j_{2}i} \mbox{\boldmath $\xi$}_i^{0,k}t &
\cdots & \mbox{\boldmath$\eta$}_{j_k}^{k} + \sum_{i=1}^N
\omega_{j_{k}i} \mbox{\boldmath $\xi$}_i^{0,k}t
\end{array} \right| ,
\label{eq:delta_j1}
\end{equation}
with $1\leq j_1 < j_2 < \ldots < j_k \leq N$. Using standard
properties of determinants, we easily see that the time evolution
of $\Delta_{j_1,j_2,\ldots,j_{k}}^{0,k}$ is mainly determined by
the behavior of determinants of the form
\begin{equation}
\left| \begin{array}{cccc} \omega_{j_{1}m_{1}} \mbox{\boldmath
$\xi$}_{m_{1}}^{0,k}t & \omega_{j_{2}m_{2}} \mbox{\boldmath
$\xi$}_{m_{2}}^{0,k}t & \cdots & \omega_{j_{k}m_{k}}
\mbox{\boldmath $\xi$}_{m_{k}}^{0,k}t
\end{array} \right| =
t^k \prod_{i=1}^{k} \omega_{j_i m_i}\cdot \left|
\begin{array}{cccc}  \mbox{\boldmath
$\xi$}_{m_{1}}^{0,k} &  \mbox{\boldmath $\xi$}_{m_{2}}^{0,k} &
\cdots &  \mbox{\boldmath $\xi$}_{m_{k}}^{0,k}
\end{array} \right| \propto t^k,
\label{eq:orizousa_1}
\end{equation}
where $m_i \in \{1,2,\ldots,N\}$, $i=1,2,\ldots,k$, with $m_i\neq
m_j$, for all $i\neq j$. Thus, from (\ref{eq:M_prop_tk}) and
(\ref{eq:orizousa_1}) we conclude that the contribution to the
behavior of $S_k'$ of the determinants related to
$\Delta_{j_1,j_2,\ldots,j_{k}}^{0,k}$ is to provide constant terms
in (\ref{eq:ord_sk}). All other determinants appearing in the
definition of $S_k'$, not being of the form of
$\Delta_{j_1,j_2,\ldots,j_{k}}^{0,k}$, contain at least one column
from the first $N$ columns of matrix $\textbf{D}^{0,k}$ and
introduce in (\ref{eq:ord_sk}) terms that grow at a rate {\it
slower} than $t^{k}$, which will ultimately have no bearing on the
behavior of GALI$_k$(t). To see this, let us consider a particular
determinant of this kind
\begin{equation}
\Delta_m^{0,k}=\left|
\begin{array}{cccccc}
\mbox{\boldmath $\xi$}_1^{0,k} & \cdots & \mbox{\boldmath
$\xi$}_m^{0,k} & \mbox{\boldmath$\eta$}_{1}^{k} + \sum_{i=1}^N
\omega_{1i} \mbox{\boldmath $\xi$}_i^{0,k}t & \cdots &
\mbox{\boldmath$\eta$}_{k-m}^{k} + \sum_{i=1}^N \omega_{k-m\, i}
\mbox{\boldmath $\xi$}_{i}^{0,k}t
\end{array} \right|,
\label{eq:delta_1}
\end{equation}
containing the first $m$ columns of matrix $\textbf{D}^{0,k}$,
which are related to the action coordinates of the system, and the
first $k-m$ columns of the angle related columns of
$\textbf{D}^{0,k}$, with $1\leq m\leq k$. The first $m$ columns of
$\Delta_m^{0,k}$ are time independent. Using repeatedly a standard
property of determinants, we easily see that the time evolution of
$\Delta_m^{0,k}$ is mainly determined by the time evolution of
determinants of the form:
\begin{equation}
\left| \begin{array}{cccccccc} \mbox{\boldmath $\xi$}_1^{0,k} &
\mbox{\boldmath $\xi$}_2^{0,k} &\cdots & \mbox{\boldmath
$\xi$}_m^{0,k} & \omega_{1i_1} \mbox{\boldmath $\xi$}_{i_1}^{0,k}t
&\omega_{2i_2} \mbox{\boldmath $\xi$}_{i_2}^{0,k}t&\cdots
&\omega_{k-m \, i_{k-m}} \mbox{\boldmath $\xi$}_{i_{k-m}}^{0,k}t
\end{array} \right| \propto t^{k-m},
\label{eq:orizousa_2}
\end{equation}
with $i_j \in \{m+1, m+2, \ldots, N \}$, $j=1,2,\ldots, k-m$ and
$i_j \neq i_l$, for all $j\neq l$. Thus, the contribution to the
behavior of $S_k'$ of  determinants similar to $\Delta_{m}^{0,k}$
are terms proportional to $t^{k-m}/t^k = 1/t^m$ ($1\leq m \leq k$),
tending to zero as $t$ grows. Since the $k \times k$ determinants
appearing in the definition of $S_k'$ involve both terms of the
form (\ref{eq:delta_j1}), growing as $t^k$ and of the form
(\ref{eq:delta_1}), growing as $t^{k-m}$, the overall behavior of
$S_k'$ will be defined by determinants growing as $t^k$, which
when combined with (\ref{eq:M_prop_tk}) yields the important
result
\begin{equation}
\mbox{GALI}_k (t) \approx \mbox{constant} \,\,\, \mbox{for} \,\,\,
2 \leq k \leq N. \label{eq:GALI_order_1}
\end{equation}

Next, let us now turn to the case of $k$ deviation vectors with $N
< k \leq 2N$. The fastest growing determinants are again those
containing the last $N$ columns of the matrix $\textbf{D}^{0,k}$:
\begin{equation}
\Delta_{j_1,j_2,\ldots,j_{k-N},1,2,\ldots,N}^{0,k}= \left|
\begin{array}{cccccc}
\mbox{\boldmath $\xi$}_{j_1}^{0,k} & \cdots & \mbox{\boldmath
$\xi$}_{j_{k-N}}^{0,k} & \mbox{\boldmath$\eta$}_{1}^{k} +
\sum_{i=1}^N \omega_{1i} \mbox{\boldmath $\xi$}_i^{0,k}t & \cdots
& \mbox{\boldmath$\eta$}_{N}^{k} + \sum_{i=1}^N \omega_{N i}
\mbox{\boldmath $\xi$}_{i}^{0,k}t
\end{array} \right|,
\label{eq:delta_2}
\end{equation}
with $1\leq j_1 < j_2 < \ldots < j_{k-N} \leq N$. The first $k-N$
columns of $\Delta_{j_1,j_2,\ldots,j_{k-N},1,2,\ldots,N}^{0,k}$
are chosen among the first $N$  columns of $\textbf{D}^{0,k}$
which are time independent. So there exist $N!/((k-N)!(2N-k)!)$
determinants of the form (\ref{eq:delta_2}), which can be written
as a sum of simpler $k \times k$ determinants, each containing in
the position of its last $N$ columns $\mbox{\boldmath$\eta$}_i^k$,
$i=1,2,\ldots,N$ and/or columns of the form
$\omega_{ji}\mbox{\boldmath $\xi$}_i^{0,k}t$ with
$i,j=1,2,\ldots,N$. We exclude the ones where $\mbox{\boldmath
$\xi$}_i^{0,k}$, $i=1,2,\ldots,N$ appear more than once, since in
that case the corresponding determinant is zero. Among the
remaining determinants, the fastest increasing ones are those
containing as many columns proportional to $t$ as possible.

Since $t$ is always multiplied by the $\mbox{\boldmath
$\xi$}_i^{0,k}$, and such columns occupy the first $k-N$ columns
of $\Delta_{j_1,j_2,\ldots,j_{k-N},1,2,\ldots,N}^{0,k}$, $t$
appears at most $N-(k-N)=2N-k$ times. Otherwise the determinant
would contain the same $\mbox{\boldmath $\xi$}_i^{0,k}$ column at
least twice and would be equal to zero. The remaining
$k-(2N-k)-(k-N)=k-N$ columns are filled by the $\mbox{\boldmath
$\eta$}_i^{k}$ each of which appears at most once. Thus, the time
evolution of $\Delta_{j_1,j_2,\ldots,j_{k-N},1,2,\ldots,N}^{0,k}$
is mainly determined by determinants of the form:
\begin{equation}
\left| \begin{array}{ccccccccc} \mbox{\boldmath $\xi$}_{j_1}^{0,k}
& \cdots & \mbox{\boldmath $\xi$}_{j_{k-N}}^{0,k} &
\mbox{\boldmath $\eta$}_{i_1}^{0,k} & \cdots & \mbox{\boldmath
$\eta$}_{i_{k-N}}^{0,k} &\omega_{i_{k-N+1}m_1} \mbox{\boldmath
$\xi$}_{m_1}^{0,k}t &\cdots &\omega_{i_{N} m_{2N-k}}
\mbox{\boldmath $\xi$}_{i_{2N-k}}^{0,k}t
\end{array} \right| \propto t^{2N-k},
\label{eq:orizousa_delta_2}
\end{equation}
with $i_l  \in \{1,2,\ldots,N\}$, $l=1,2,\ldots,N$, $i_l \neq
i_j$, for all $l\neq j$ and $m_l  \in \{1,2,\ldots,N\}$,
$l=1,2,\ldots,2N-k$, $m_l \not\in \{j_1,j_2,\ldots, j_{k-N} \}$,
$m_l \neq m_j$, for all $l\neq j$. So determinants of the form
(\ref{eq:delta_2}) contribute to the time evolution of $S_k'$ by
introducing terms proportional to $t^{2N-k}/t^k=1/t^{2(k-N)}$. All
other determinants appearing in the definition of $S_k'$, not
having the form of
$\Delta_{j_1,j_2,\ldots,j_{k-N},1,2,\ldots,N}^{0,k}$, introduce
terms that tend to zero faster than $1/t^{2(k-N)}$ since they
contain more than $k-N$ time independent columns of the form
$\mbox{\boldmath $\xi$}_i^{0,k}$, $i=1,2,\ldots,N$. Thus $S_k'$
and consequently GALI$_k$ tend to zero following a power law of
the form:
\begin{equation}
\mbox{GALI}_k (t) \propto \frac{1}{t^{2(k-N)}}\,\,\, \mbox{for}
\,\,\, N < k \leq 2N. \label{eq:GALI_order_2}
\end{equation}

\subsubsection{The case of $m>0$ tangent initial deviation vectors}
\label{with tangent vectors}

Finally, let us consider the behavior of GALI$_k$ for the special
case where $m$ initial deviation vectors, with $m\leq k$ and
$m\leq N$, are located in the tangent space of the torus. In this
case, matrix $\textbf{D}$, whose elements appear in the definition
of $S_k'$, has the form given by (\ref{eq:matix _D2a}). Thus, all
determinants appearing in the definition of $S_k'$ have as a
common factor the quantity $1/ \prod_{i=1}^{k-m} M_{m+i}(t)$,
which decreases to zero following a power law
\begin{equation}
\frac{1}{\prod_{i=1}^{k-m} M_{m+i}(t)} \propto \frac{1}{t^{k-m}}
.\label{eq:M_prop_tkm}
\end{equation}
Proceeding in exactly the same manner as in the $m=0$ case above,
we deduce that, in the case of $2\leq k \leq N$ the fastest
growing $k\times k$ determinants resulting from the matrix
$\textbf{D}^{m,k}$ are of the form:
\begin{equation}
\left| \begin{array}{cccccccc} \mbox{\boldmath $\eta$}_{i_{1}}^{k}
& \mbox{\boldmath $\eta$}_{i_{2}}^{k} & \cdots & \mbox{\boldmath
$\eta$}_{i_{m}}^{k} & \omega_{i_{m+1}n_{1}} \mbox{\boldmath
$\xi$}_{n_{1}}^{0,k}t & \omega_{i_{m+2}n_{2}} \mbox{\boldmath
$\xi$}_{n_{2}}^{0,k}t & \cdots & \omega_{i_{k}n_{k-m}}
\mbox{\boldmath $\xi$}_{n_{k-m}}^{0,k}t
\end{array} \right|  \propto t^{k-m},
\label{eq:orizousa_m1}
\end{equation}
with $i_l \in \{1,2,\ldots,N\}$, $l=1,2,\ldots,k$ with $i_l\neq
i_j$ for $l\neq j$, and $n_l \in \{1,2,\ldots,N\}$,
$l=1,2,\ldots,k-m$ with $n_l\neq n_j$, for $l\neq j$. Hence, we
conclude that the behavior of $S_k'$, and consequently of GALI$_k$
is defined by the behavior of determinants having the form of
(\ref{eq:orizousa_m1}) which, when combined with
(\ref{eq:M_prop_tkm}) implies that
\begin{equation}
\mbox{GALI}_k (t) \approx \mbox{constant} \,\,\, \mbox{for} \,\,\,
2 \leq k \leq N. \label{eq:GALI_order_3}
\end{equation}

The case of $N < k \leq 2N$ deviation vectors, however, with $m>0$
initially tangent vectors, yields a considerably different result.
Following entirely analogous arguments as in the $m=0$ case, we
find that, if $m<k-N$, $S_k'$ and GALI$_k$ evolve proportionally
to $t^{2N-k}/t^{k-m}= 1/t^{2(k-N)-m}$. On the other hand, if
$m\geq k-N$, one can show that the fastest growing determinant is
proportional to $t^{N-m}$. In this case, $S_k'$ and GALI$_k$
evolve in time following a quite different power law:
$t^{N-m}/t^{k-m}=1/t^{k-N}$.

Summarizing the results of this section, we see that GALI$_k$ for
regular motion remains essentially constant when $k\leq N$, while
it tends to zero for $k> N$ following a power law which depends on
the number $m$ ($m\leq N$ and $m \leq k$) of deviation vectors
initially tangent to the torus. In conclusion, we have shown that:
\begin{equation}
\mbox{GALI}_k (t) \propto \left\{ \begin{array}{ll}
\mbox{constant} &
\mbox{if $2\leq k \leq N$} \\
\frac{1}{t^{2(k-N)-m}} & \mbox{if $N< k \leq 2N$ and $0\leq m <
k-N$} \\
\frac{1}{t^{k-N}} & \mbox{if $N< k \leq 2N$ and $m \geq
k-N$} \\
\end{array} \right.
. \label{eq:GALI_order_all}
\end{equation}

\section{Numerical verification and applications}
\label{GALI_a}

In order to apply the GALI method to Hamiltonian systems and
verify  the theoretically predicted behavior of the previous
sections, we shall use two simple examples with 2 (2D) and 3 (3D)
degrees of freedom: the well--known 2D H\'{e}non--Heiles system
\cite{HH64}, described by the Hamiltonian
\begin{equation}
H_2 = \frac{1}{2} (p_x^2+p_y^2) + \frac{1}{2} (x^2+y^2) + x^2 y -
\frac{1}{3} y^3, \label{eq:2DHam}
\end{equation}
and the 3D Hamiltonian system:
\begin{equation}
H_3 = \sum_{i=1}^3 \frac{\omega_i}{2} (q_i^2+p_i^2) + q_1^2
q_2+q_1^2 q_3, \label{eq:3DHam}
\end{equation}
studied in \cite{CGG78,BGGS80b}. We keep the parameters of the two
systems fixed at the energies $H_2=0.125$ and $H_3=0.09$, with
$\omega_1=1$, $\omega_2=\sqrt{2}$ and $\omega_3=\sqrt{3}$. In
order to illustrate the behavior of GALI$_k$, for different values
of $k$, we shall consider some representative cases of chaotic and
regular orbits of the two systems.

Additionally, we shall study the higher--dimensional example of a
15D Hamiltonian, describing a chain of 15 particles with quadratic
and quartic nearest neighbor interaction, known as the famous
Fermi--Pasta--Ulam (FPU) model \cite{FPU55}
\begin{equation}
H_{15} = \frac{1}{2} \sum_{i=1}^{15} p_i^2 + \sum_{i=1}^{15}
\left[\frac{1}{2}(q_{i+1}-q_i)^2 + \frac{1}{4}\beta
(q_{i+1}-q_i)^4 \right] \label{eq:15DHam}
\end{equation}
where $q_i$ is the displacement of the $i$th particle from its
equilibrium point and $p_i$ is the conjugate momentum. This is a
model we have recently analyzed in \cite{ABS05} and we shall use
here the same values of the energy $H_{15}=26.68777$ and
$\beta=1.04$ as in that study.

\subsection{A 2D Hamiltonian system}
\label{App_2D}

\begin{figure}\vspace{1cm}
\centerline{\includegraphics[width=7.5cm,height=7.5cm]{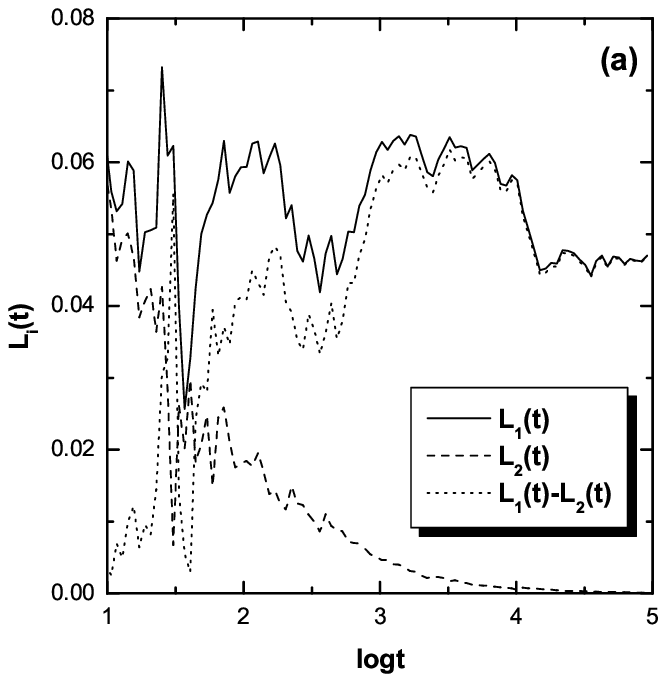}
\includegraphics[width=7.5cm,height=7.5cm] {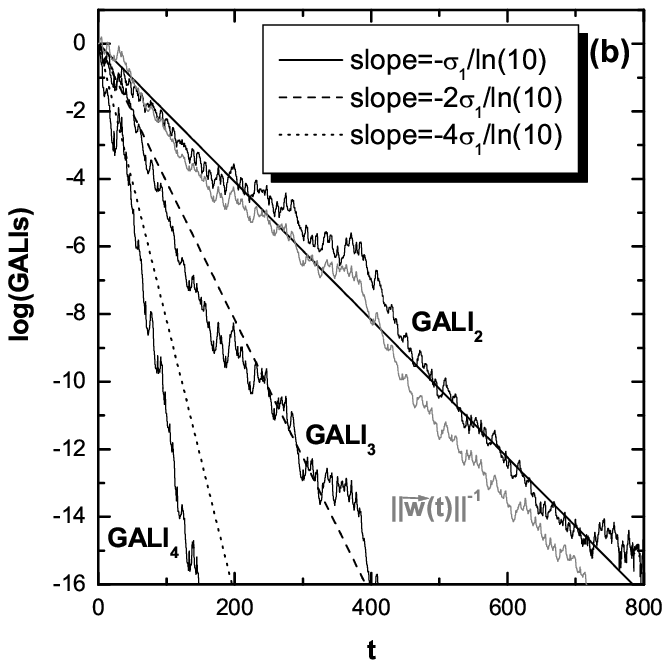} }
\caption{(a) The evolution of $L_1(t)$ (solid curve), $L_2(t)$
(dashed curve) and $L_1(t)-L_2(t)$ (dotted curve) for a chaotic
orbit with initial conditions $x=0$, $y=-0.25$, $p_x=0.42$,
$p_y=0$ of the 2D system (\ref{eq:2DHam}). (b) The evolution of
GALI$_2$, GALI$_3$ and GALI$_4$ of the same orbit. The plotted
lines correspond to functions proportional to $e^{-\sigma_1 t}$
(solid line), $e^{-2\sigma_1 t}$ (dashed line) and $e^{-4\sigma_1
t}$ (dotted line) for $\sigma_1=0.047$. Note that the $t$--axis is
linear. The evolution of the norm of the deviation vector $\vec{w}(t)$
(with $\|\vec{w}(0)\|=1$) used for the computation of $L_1(t)$, is
also plotted in (b) (gray curve).} \label{fig:2D_ch_L1}
\end{figure}

Let us consider first a chaotic orbit of the 2D Hamiltonian
(\ref{eq:2DHam}), with initial conditions $x=0$, $y=-0.25$,
$p_x=0.42$, $p_y=0$. In figure \ref{fig:2D_ch_L1}(a)
we see the time evolution of $L_1(t)$ of this orbit. The
computation is carried out until $L_1(t)$ stops having large
fluctuations and approaches a positive value (indicating the
chaotic nature of the orbit), which could be considered as a good
approximation of the maximal LCE, $\sigma_1$. Actually, for
$t\approx 10^5$, we find $\sigma_1\approx 0.047$.

We recall that 2D Hamiltonian systems have only one positive LCE
$\sigma_1$, since the second largest is $\sigma_2=0$. It also
holds that $\sigma_3=-\sigma_2$ and $\sigma_4=-\sigma_1$ and thus
formula (\ref{eq:ch_GALI_decay}), which describes the time
evolution of GALI$_k$ for chaotic orbits, gives
\begin{equation}
\mbox{GALI}_2(t) \propto e^{-\sigma_1 t},\,\,\, \mbox{GALI}_3(t)
\propto e^{-2\sigma_1 t},\,\,\,\mbox{GALI}_4(t) \propto
e^{-4\sigma_1 t}. \label{eq:2DHam_ch_GALIs}
\end{equation}
In figure \ref{fig:2D_ch_L1}(b) we plot GALI$_k$, $k=2,3,4$ for
the same chaotic orbit as a function of time $t$. We plot $t$ in
linear scale so that, if (\ref{eq:2DHam_ch_GALIs}) is valid, the
slope of GALI$_2$, GALI$_3$ and GALI$_4$ should approximately be
$-\sigma_1/ \ln 10$, $-2\sigma_1/ \ln 10$ and $-4\sigma_1/ \ln 10$
respectively. From figure \ref{fig:2D_ch_L1}(b) we see that lines
having precisely these slopes, for $\sigma_1=0.047$, approximate
quite accurately the computed values of the GALIs. The biggest
deviation between the theoretical curve and  numerical data
appears in the case of GALI$_4$ where the theoretical prediction
underestimates the decaying rate of the index, but even in this
case the difference does not appear too significant. Note,
however, the important difference in the times it takes to decide
about the chaotic nature of the orbit: Waiting for the maximal LCE
to converge in figure \ref{fig:2D_ch_L1}(a), one needs more than
$10^4$ time units, while, as we see in figure
\ref{fig:2D_ch_L1}(b), the GALI$_k$'s provide this information in
less than $400$ time units!

We also note that, plotting in this example the evolution of the
quantity $\|\vec{w}(t)\|^{-1}$ (with $\|\vec{w}(0)\|=1$), which is
used to determine $L_1(t)$ in (\ref{eq:lyap1_def}) and is practically
identified with the Fast Lyapunov Indicator (FLI), we obtain in figure
\ref{fig:2D_ch_L1}(b) a graph similar to that of GALI$_2(t)$.  This is
not surprising, as both $\|\vec{w}(t)\|^{-1}$ and GALI$_2(t)$ tend
exponentially to zero following a decay proportional to $e^{-\sigma_1
t}$ (see equations (\ref{eq:vector_norm}) and
(\ref{eq:2DHam_ch_GALIs})).  From the results of figure
\ref{fig:2D_ch_L1}(b) we see that the different plotted quantities
reach the limit of computer's accuracy ($10^{-16}$) at different times
and in particular GALI$_2$ at $t\approx 800$, GALI$_3$ at $t\approx
400$, GALI$_4$ at $t\approx 150$ and $\|\vec{w}(t)\|^{-1}$ at
$t\approx 720$.  The CPU time needed for computing the evolution of
the indices up to these times were: 0.220 sec for
$\|\vec{w}(t)\|^{-1}$, 0.295 sec for GALI$_2$, 0.165 sec for GALI$_3$
and 0.070 sec for GALI$_4$ respectively. Thus, in this case also, it
is clear that the higher order GALI$_k$ (with $k>2$) can identify the
chaotic nature of an orbit faster than the methods of the maximal LCE,
the FLI or the SALI (equivalent to GALI$_2$, see below).

It is interesting to remark at this point (as mentioned in section
\ref{GALI_chaos}), that the accuracy of the exponential laws
(\ref{eq:2DHam_ch_GALIs}) is due to the fact that the local
Lyapunov exponents cease to fluctuate significantly about their
limit values, after a relatively short time interval. To see this,
we have plotted in figure \ref{fig:2D_ch_L1}(a), the two nonnegative
local Lyapunov exponents $L_1(t)$, $L_2(t)$, as well as their
difference. Note that $L_1(t)-L_2(t)$ begins to be well
approximated by $\sigma_1-\sigma_2=\sigma_1$ already for times $t$ of order
$10^2$ units. A similar behavior of such $L_1(t)-L_{i}(t)$,
$i=2,3,\ldots,2N$ differences are observed for the other
Hamiltonians we studied in this paper having 3 or more degrees of
freedom.

\begin{figure}\vspace{1cm}
\centerline{\includegraphics[width=7.5cm,height=7.5cm] {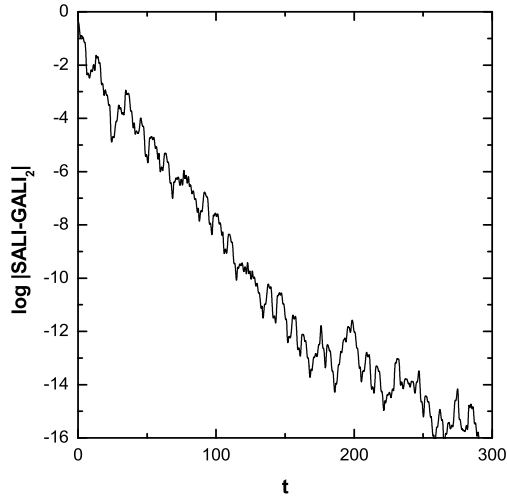}}
\caption{The absolute difference between GALI$_2$ and SALI of the
chaotic orbit of figure \ref{fig:2D_ch_L1} as a function of time
$t$.} \label{fig:2D_ch_GALI2_SALI}
\end{figure}

As explained in detail in Appendix \ref{GALI_2}, GALI$_2$ practically
coincides with SALI in the case of chaotic orbits. This becomes
evident from figure \ref{fig:2D_ch_GALI2_SALI}
where we plot the absolute difference between GALI$_2$ and SALI of
the chaotic orbit of figure \ref{fig:2D_ch_L1} as a function of
time $t$. The two indices practically coincide after about $t
\approx 300$ units, since their difference is at the limit of
computer's accuracy ($10^{-16}$), although their actual values are
of order $10^{-5}$ (see figure \ref{fig:2D_ch_L1}(b)).

Let us now study the behavior of GALI$_k$ for a regular orbit of
the 2D Hamiltonian (\ref{eq:2DHam}). From
(\ref{eq:GALI_order_all}) it follows that in the case of a
Hamiltonian system with $N=2$ degrees of freedom GALI$_2$ will
always remain different from zero, while GALI$_3$ and GALI$_4$
should decay to zero following a power law, whose exponent depends
on the number $m$ of deviation vectors that are initially tangent
to the torus on which the orbit lies. Now, for a regular orbit of
the 2D Hamiltonian (\ref{eq:2DHam}) and a random choice of initial
deviation vectors, we expect the GALI indices to behave as
\begin{equation}
\mbox{GALI}_2(t) \propto \mbox{constant},\,\,\, \mbox{GALI}_3(t)
\propto \frac{1}{t^2},\,\,\,\mbox{GALI}_4(t) \propto
\frac{1}{t^4}. \label{eq:2DHam_or_GALIs}
\end{equation}

\begin{figure}\vspace{1cm}
\centerline{\includegraphics[width=7.5cm,height=7.5cm]{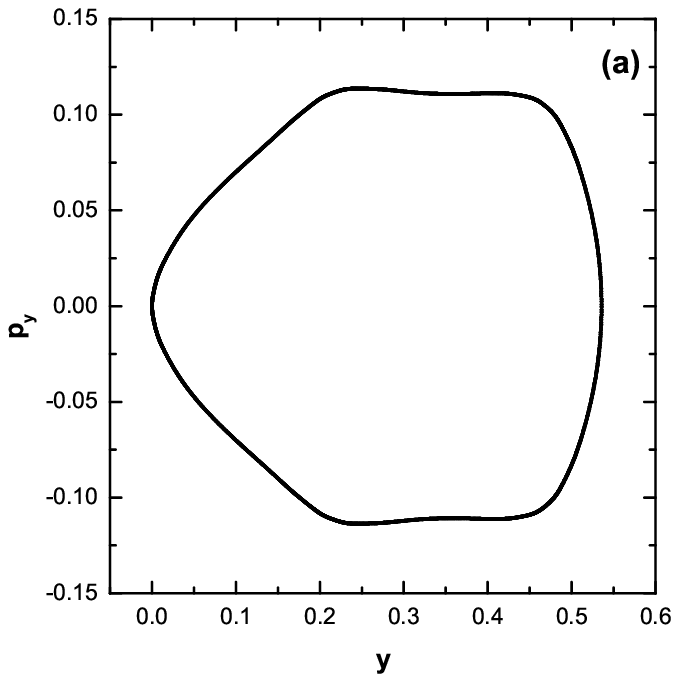}
\includegraphics[width=7.5 cm,height=7.5 cm] {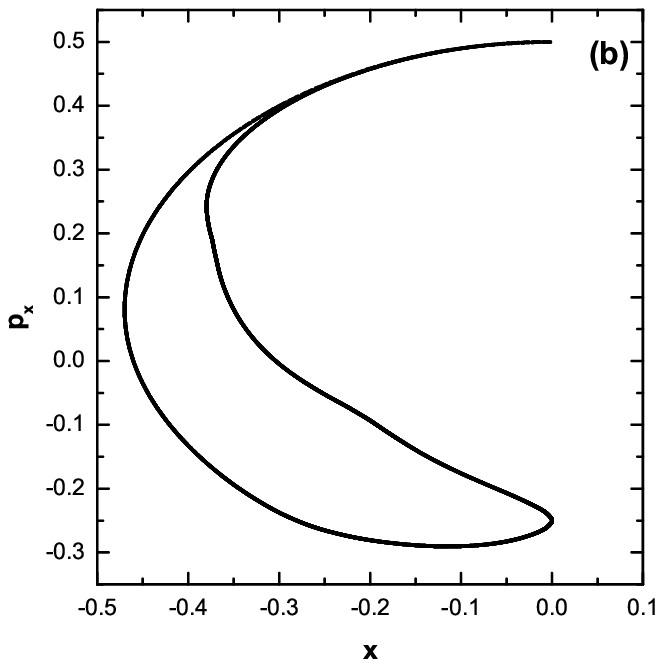} }
\caption{The Poincar\'{e} Surface of Section (PSS) defined by (a)
$x=0$ and (b) $y=0$ of the regular orbit with initial conditions
$x=0$, $y=0$, $p_x=0.5$, $p_y=0$ for the H\'{e}non--Heiles system
(\ref{eq:2DHam}).} \label{fig:2D_or_PSS}
\end{figure}

A simple qualitative way of studying the dynamics of a Hamiltonian
system is by plotting the successive intersections of the orbits
with a Poincar\'{e} Surface of Section (PSS) \cite{LL92}. In 2D
Hamiltonians, the PSS is a two dimensional plane and the points of
a regular orbit (which lie on a torus) fall on a smooth closed
curve. This property allows us to choose initial deviation vectors
tangent to a torus in the case of system (\ref{eq:2DHam}). In
particular, we consider the regular orbit with initial conditions
$x=0$, $y=0$, $p_x=0.5$, $p_y=0$. In figure \ref{fig:2D_or_PSS},
we plot the intersection points of this orbit with the PSS defined
by $x=0$ (panel (a)) and $y=0$ (panel (b)). From the morphology of
the two closed curves of figure \ref{fig:2D_or_PSS}, it is easily
seen that deviation vectors $\hat{e}_1=(1,0,0,0)$ and
$\hat{e}_4=(0,0,0,1)$ are tangent to the torus.

\begin{figure}
\centerline{
\begin{tabular}{cc}
\includegraphics[width=7.5cm,height=7.5cm]{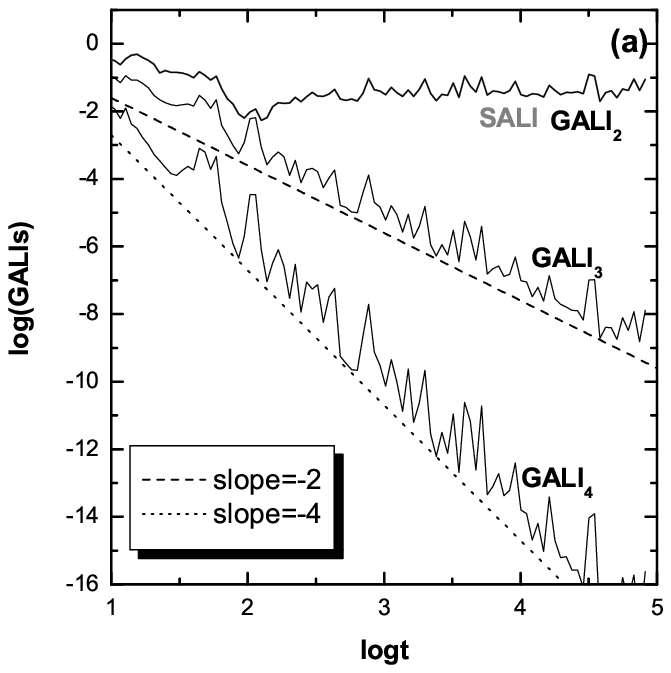} &
\includegraphics[width=7.5cm,height=7.5cm]{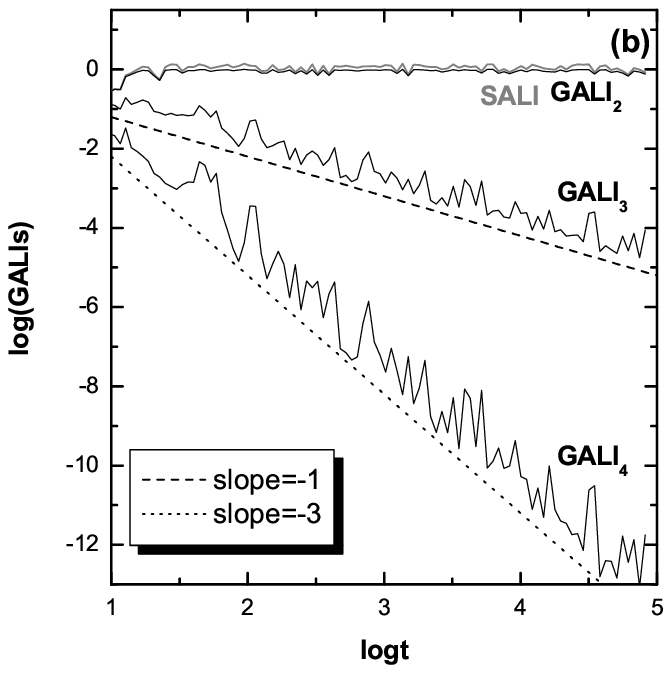} \\
\multicolumn{2}{c}{\includegraphics[width=7.5cm,height=7.5cm]{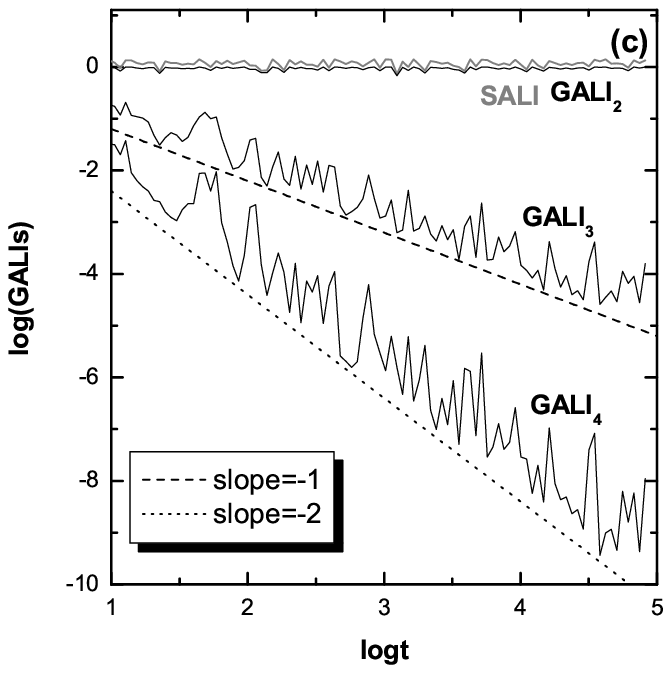}}
\end{tabular}}
\caption{Time evolution of SALI (gray curves), GALI$_2$, GALI$_3$
and GALI$_4$ for the regular orbit of figure \ref{fig:2D_or_PSS}
in log--log scale for different values of the number $m$ of
deviation vectors initially tangent to the torus: (a) $m=0$, (b)
$m=1$ and (c) $m=2$. We note that in panel (a) the curves of SALI
and GALI$_2$ are very close to each other and thus cannot be
distinguished. In every panel, dashed lines corresponding to
particular power laws are also plotted.} \label{fig:2D_or_GALIs}
\end{figure}

In Figure \ref{fig:2D_or_GALIs}, we plot the time evolution of
SALI, GALI$_2$, GALI$_3$ and GALI$_4$ for the regular orbit of
figure \ref{fig:2D_or_PSS}, for various choices of initial
deviation vectors. In figure \ref{fig:2D_or_GALIs}(a) the initial
deviation vectors are randomly chosen so that none of them is
tangent to the torus. In this case SALI and GALI$_2$ fluctuate
around non--zero values, while GALI$_3$ and GALI$_4$ tend to zero
following the theoretically  predicted power laws, see
(\ref{eq:2DHam_or_GALIs}). In figure \ref{fig:2D_or_GALIs}(b) we
present results for the indices when we have $m=1$ initial
deviation vector tangent to the torus (in particular vector
$\hat{e}_1$). In this case the indices evolve as predicted by
(\ref{eq:GALI_order_all}), i.~e.~SALI and GALI$_2$ remain
practically constant, while $\mbox{GALI}_3 \propto 1/t$ and
$\mbox{GALI}_4 \propto 1/t^3$. Finally, in figure
\ref{fig:2D_or_GALIs}(c) we have plotted our results using $m=2$
initial deviation vectors tangent to the torus (vectors
$\hat{e}_1$ and $\hat{e}_4$). Again the predictions of
(\ref{eq:GALI_order_all}) are seen to be valid since
$\mbox{GALI}_3 \propto 1/t$ and $\mbox{GALI}_4 \propto 1/t^2$.

The different behavior of SALI (or GALI$_2$) for regular and
chaotic orbits has already been successfully used for
discriminating between regions of order and chaos in various
dynamical systems \cite{SABV04,SESS04,MA05b,MA05a,MA06,BS05,CDLMV06}. For
example, by integrating orbits whose initial conditions lie on a
grid, and by attributing to each grid point a color according to
the value of SALI at the end of a given integration time, one can
obtain clear and informative pictures of the dynamics in the full
phase space of several Hamiltonian systems of physical
significance \cite{SABV04,SESS04,BS05}.

Figures \ref{fig:2D_ch_L1}(b) and \ref{fig:2D_or_GALIs} clearly
illustrate that GALI$_3$ and GALI$_4$ tend to zero both for
regular and chaotic orbits, but with very different time rates. We
may use this difference to distinguish between chaotic and regular
motion following a different approach than SALI or GALI$_2$. Let
us illustrate this by considering the computation of GALI$_4$:
From (\ref{eq:2DHam_ch_GALIs}) and (\ref{eq:2DHam_or_GALIs}), we
expect $\mbox{GALI}_4 \propto e^{-4 \sigma_1 t}$ for chaotic
orbits and $\mbox{GALI}_4 \propto 1/t^4$ for regular ones. These
time rates imply that, in general, the time needed for the index
to become zero is much larger for regular orbits. Thus, instead of
simply registering the value of the index at the end of a given
time interval (as we do with SALI or GALI$_2$), let us record the
time, $t_{th}$, needed for GALI$_4$ to reach a very small
threshold, e.~g.~$10^{-12}$, and color each grid point according
to the value of $t_{th}$.

\begin{figure}\vspace{0cm}
\centerline{\includegraphics[width=7.5 cm,height=7.5 cm] {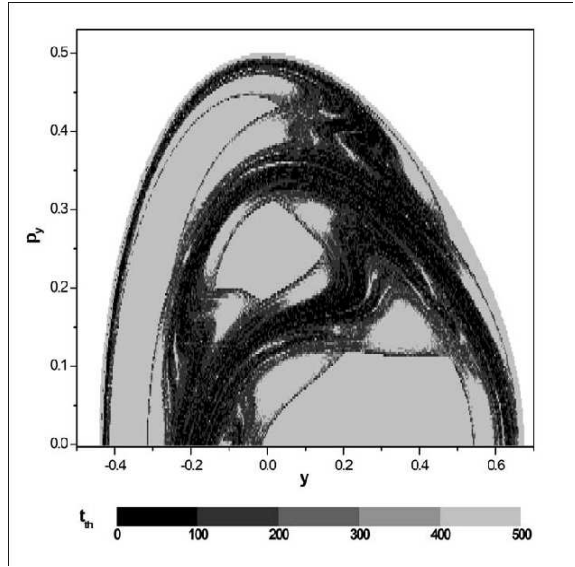}}
\vspace{0cm} \caption{Regions of different values of the time
$t_{th}$ needed for GALI$_4$ to become less than $10^{-12}$ on the
PSS defined by $x=0$ of the 2D H\'{e}non--Heiles Hamiltonian
(\ref{eq:2DHam}).} \label{fig:2D_scan}
\end{figure}

The outcome of this procedure for the 2D H\'{e}non--Heiles system
(\ref{eq:2DHam}) is presented in figure \ref{fig:2D_scan}.
Each orbit is integrated up to $t=500$ units and if the value of
GALI$_4$, at the end of the integration is larger than $10^{-12}$
the corresponding grid point is colored by  the light gray color
used for $t_{th} \geq 400$. Thus we can clearly distinguish in
this figure among various `degrees' of chaotic behavior in regions
colored black or dark gray -- corresponding to small values of
$t_{th}$ -- and regions of regular motion colored light gray,
corresponding to large values of $t_{th}$. At the border between
them we find points having intermediate values of $t_{th}$ which
belong to the so--called `sticky' chaotic regions. Thus, this
approach yields a very detailed chart of the dynamics, where even
tiny islands of stability can be identified inside the large
chaotic sea. We note that for every initial condition the same set
of initial deviation vectors was used, ensuring the same initial
value of GALI$_4$ for all orbits and justifying the dynamical
interpretation of the color scale of figure \ref{fig:2D_scan}.

\subsection{A 3D Hamiltonian system}
\label{App_3D}

Let us now study the behavior of the GALIs in the case of the 3D
Hamiltonian (\ref{eq:3DHam}). Following \cite{CGG78,BGGS80b} the
initial conditions of the orbits of this system are defined by
assigning arbitrary values to the positions $q_1$, $q_2$, $q_3$,
as well as the so--called `harmonic energies' $E_1$, $E_2$, $E_3$
related to the momenta through
\begin{equation}
p_i=\sqrt{\frac{2 E_i}{\omega_i}}\,\,\, , \,\,\,i=1,2,3.
\label{eq:harm_en}
\end{equation}
Chaotic orbits of 3D Hamiltonian systems generally have two
positive Lyapunov exponents, $\sigma_1$ and $\sigma_2$, while
$\sigma_3=0$. So, for approximating the behavior of GALIs
according to (\ref{eq:ch_GALI_decay}), both $\sigma_1$ and
$\sigma_2$ are needed. In particular, (\ref{eq:ch_GALI_decay})
gives
\begin{equation}  \begin{array}{c}
\mbox{GALI}_2(t) \propto e^{-(\sigma_1 - \sigma_2)t},\,\,\,
\mbox{GALI}_3(t) \propto e^{-(2 \sigma_1 - \sigma_2)t}
,\,\,\,\mbox{GALI}_4(t) \propto
e^{-(3 \sigma_1 - \sigma_2)t}, \\
\mbox{GALI}_5(t) \propto e^{-4\sigma_1 t},\,\,\, \mbox{GALI}_6(t)
\propto e^{-6\sigma_1 t}.
\end{array}\label{eq:3DHam_ch_GALIs_approx}
\end{equation}

\begin{figure}
\centerline{\includegraphics[width=7.5 cm,height=7.5 cm]{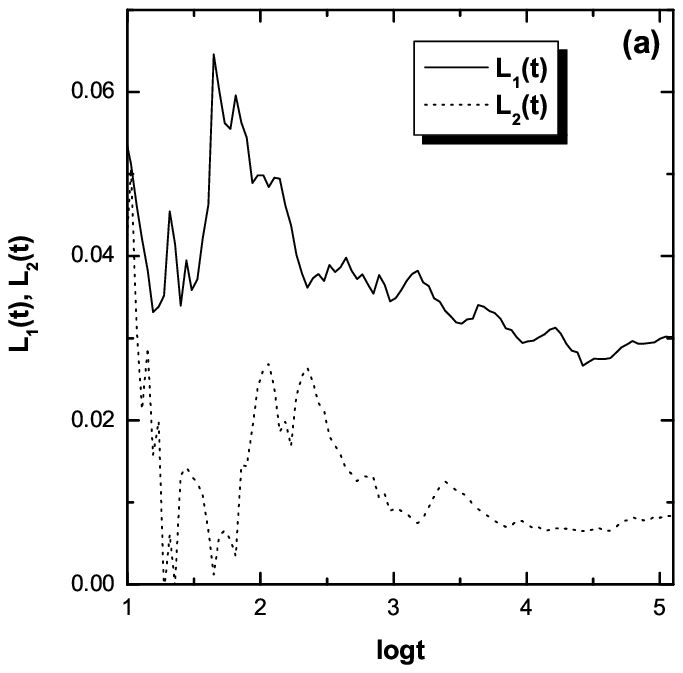}
\includegraphics[width=7.5 cm,height=7.5 cm] {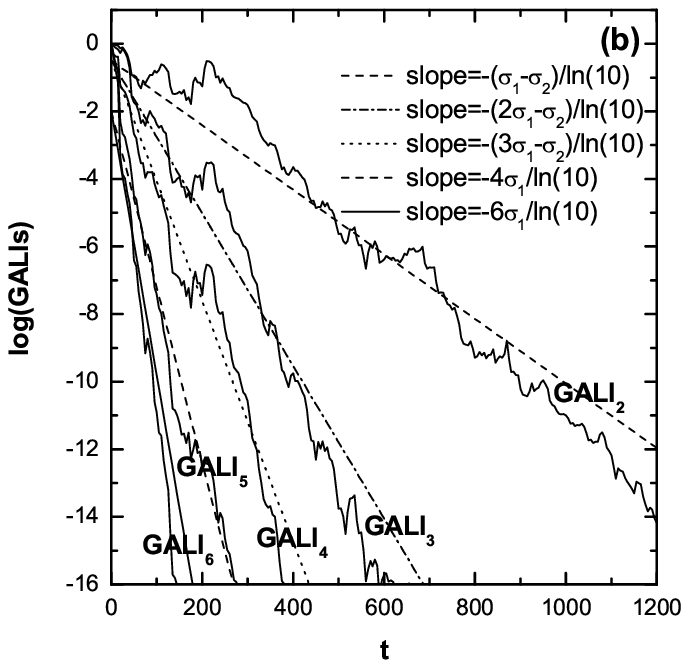} }
\caption{(a) The evolution of $L_1(t)$, $L_2(t)$ for the chaotic
orbit with initial condition $q_1=q_2=q_3=0$, $E_1=E_2=E_3=0.03$
of the 3D system (\ref{eq:3DHam}). (b) The evolution of GALI$_k$
with $k=2,\ldots,6$ of the same orbit. The plotted lines
correspond to functions proportional to $e^{-(\sigma_1 -
\sigma_2)t}$, $e^{-(2\sigma_1 - \sigma_2)t}$, $e^{-(3\sigma_1 -
\sigma_2)t}$, $e^{-4\sigma_1t}$ and $e^{-6\sigma_1t}$ for
$\sigma_1=0.03$, $\sigma_2=0.008$. Note that the $t$--axis is
linear.} \label{eq:3DHam_ch_GALIs}
\end{figure}

Let us consider the chaotic orbit with initial conditions
$q_1=q_2=q_3=0$, $E_1=E_2=E_3=0.03$ of the 3D system
(\ref{eq:3DHam}). We compute $\sigma_1$, $\sigma_2$ for this orbit
as the long time limits of the Lyapunov exponent quantities
$L_1(t)$, $L_2(t)$, applying the technique proposed by Benettin et
al.\ \cite{BGGS80b}. The results are presented in figure
\ref{eq:3DHam_ch_GALIs}(a).
The computation is carried out until $L_1(t)$ and $L_2(t)$ stop
having large fluctuations and approach some positive values (since
the orbit is chaotic), which could be considered as good
approximations of their limits $\sigma_1$, $\sigma_2$. Actually
for $t\approx 10^5$ we have $\sigma_1 \approx 0.03$ and $\sigma_2
\approx 0.008$. Using these values as good approximations of
$\sigma_1$, $\sigma_2$ we see in figure \ref{eq:3DHam_ch_GALIs}(b)
that the slopes of all GALIs are well reproduced by
(\ref{eq:3DHam_ch_GALIs_approx}).

\begin{figure}
\centerline{\includegraphics[width=7.5 cm,height=7.5 cm]{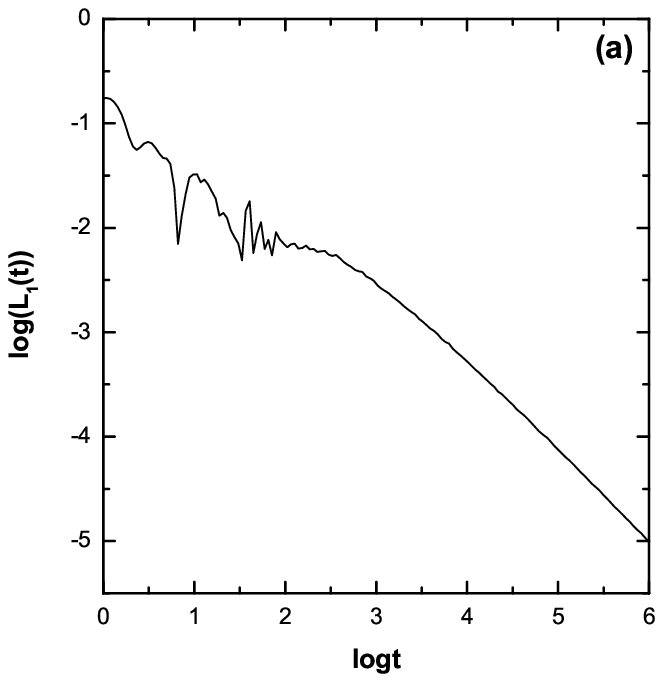}
\includegraphics[width=7.5 cm,height=7.5 cm] {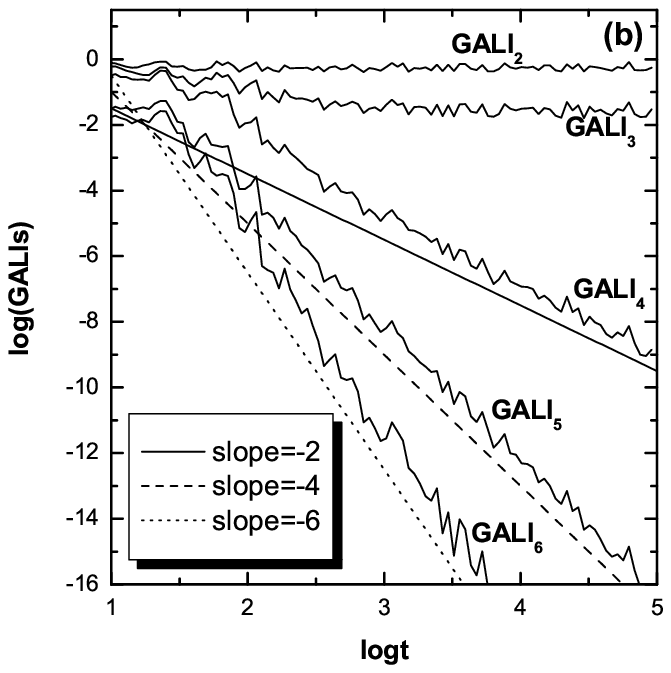} }
\caption{(a) The evolution of $L_1(t)$ for the regular orbit with
initial condition $q_1=q_2=q_3=0$, $E_1=0.005$, $E_2=0.085$,
$E_3=0$ of the 3D system (\ref{eq:3DHam}). (b) The evolution of
GALI$_k$ with $k=2,\ldots,6$ of the same orbit. The plotted lines
correspond to functions proportional to $\frac{1}{t^2}$,
$\frac{1}{t^4}$ and $\frac{1}{t^6}$.} \label{eq:3DHam_or_GALIs}
\end{figure}

Next, we consider the case of regular orbits in our 3D Hamiltonian
system. In the general case, where no initial deviation vector is
tangent to the torus where the regular orbit lies, the GALIs
should behave as:
\begin{equation} \begin{array}{c}
\mbox{GALI}_2(t) \propto \mbox{constant},\,\,\, \mbox{GALI}_3(t)
\propto \mbox{constant} ,\,\,\,\mbox{GALI}_4(t) \propto
\frac{1}{t^2}, \\
\mbox{GALI}_5(t) \propto \frac{1}{t^4},\,\,\, \mbox{GALI}_6(t)
\propto \frac{1}{t^6}.
\end{array}\label{eq:3DHam_or_GALIs_approx}
\end{equation}
according to (\ref{eq:GALI_order_all}). In order to verify
expression (\ref{eq:3DHam_or_GALIs_approx}) we shall follow a
specific regular orbit of the 3D system (\ref{eq:3DHam}) with
initial conditions $q_1=q_2=q_3=0$, $E_1=0.005$, $E_2=0.085$,
$E_3=0$. The regular nature of the orbit is revealed by the slow
convergence of its $L_1(t)$ to zero, implying that $\sigma_1=0$,
see figure \ref{eq:3DHam_or_GALIs}(a).
In figure \ref{eq:3DHam_or_GALIs}(b), we plot the values of all
GALIs of this orbit with respect to time $t$. From these results
we see that the different behaviors of GALIs are very well
approximated by formula (\ref{eq:3DHam_or_GALIs_approx}).

From the results of figures \ref{eq:3DHam_ch_GALIs} and
\ref{eq:3DHam_or_GALIs}, therefore, we conclude that in the case
of 3D Hamiltonian systems not only GALI$_2$, but also GALI$_3$ has
different behavior for regular and chaotic orbits. In particular
GALI$_3$ tends exponentially to zero for chaotic orbits (even
faster than GALI$_2$ or SALI), while it fluctuates around
non--zero values for regular orbits. Hence, the natural question
arises whether GALI$_3$ can be used instead of SALI for the faster
detection of chaotic and regular motion in 3D Hamiltonians and, by
extension, whether GALI$_k$, with $k>3$, should be preferred for
systems with $N>3$. The obvious computational drawback, of course,
is that the evaluation of GALI$_k$ requires that we numerically
follow the evolution of more than 2 deviation vectors.

First of all, let us point out that the computation of SALI,
applying (\ref{eq:SALI}), is slightly faster than GALI$_2$, for
which one needs to evaluate several $2\times 2$ determinants. For
example, for orbits of the 3D Hamiltonian (\ref{eq:3DHam}) the CPU
time needed for the computation of SALI for a fixed time interval
$t$, was about $97\%$ of the CPU time needed for the computation
of GALI$_2$ for the same time interval. Although this difference
in not significant, we prefer to compute SALI instead of GALI$_2$
and compare its efficiency with the computation of GALI$_3$.

It is obvious that the computation of GALI$_3$ for a given time
interval $t$ needs more CPU time than SALI, since we follow the
evolution of three deviation vectors instead of two. This is
particularly true for regular orbits as the index does not become
zero and its evolution has to be followed for the whole prescribed
time interval. In the case of chaotic orbits, however, the
situation is different. Let us consider, for example, the chaotic
orbit of figure \ref{eq:3DHam_ch_GALIs}. The usual technique to
characterize an orbit as chaotic is to check, after some time
interval, if its SALI has become less than a very small threshold
value, e.~g.~$10^{-8}$. For this particular orbit, this threshold
value was reached for $t\approx 760$. Adopting the same threshold
to characterize an orbit as chaotic, we find that GALI$_3$ becomes
less than $10^{-8}$ after $t\approx 335$, requiring only as much
as $65\%$ of the CPU time needed for SALI to reach the same
threshold!

\begin{figure}\vspace{0cm}
\centerline{\includegraphics[width=7.5 cm,height=7.5 cm] {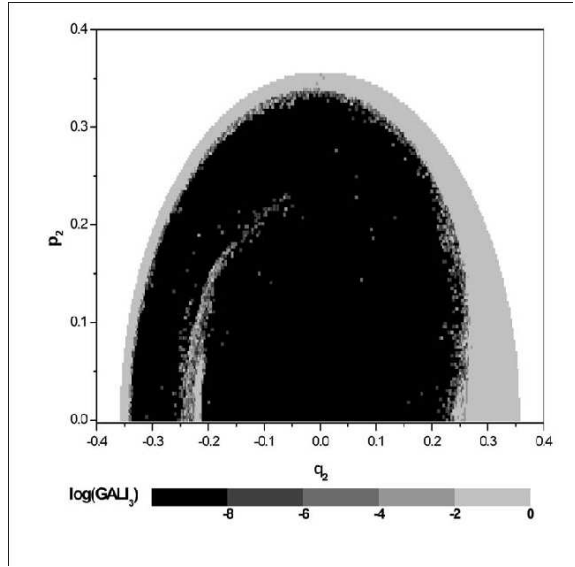}}
\vspace{0cm} \caption{Regions of different values of the GALI$_3$
on the subspace $q_3=p_3=0$, $p_2\geq 0$ of the 4--dimensional PSS
$q_1=0$ of the 3D system (\ref{eq:3DHam}) at $t=1000$.}
\label{fig:3D_scan}
\end{figure}

So, using GALI$_3$ instead of SALI, we gain considerably in CPU
time for chaotic orbits, while we lose for regular orbits. Thus,
the efficiency of using GALI$_3$ for discriminating between chaos
and order in a 3D system depends on the percentage of phase space
occupied by chaotic orbits (if all orbits are regular GALI$_3$
requires more CPU time than SALI). More crucially, however, it
depends on the choice of the final time, up to which each orbit is
integrated. As an example, let us integrate, up to $t=1000$ time
units, all orbits whose initial conditions lie on a dense grid in
the subspace $q_3=p_3=0$, $p_2\geq 0$ of a 4--dimensional PSS,
with $q_1=0$ of the 3D system (\ref{eq:3DHam}), attributing to
each grid point a color according to the value of GALI$_3$ at the
end of the integration. If GALI$_3$ of an orbit becomes less than
$10^{-8}$ for $t<1000$ the evolution of the orbit is stopped, its
GALI$_3$ value is registered and the orbit is characterized as
chaotic. The outcome of this experiment is presented in figure
\ref{fig:3D_scan}.

We find that $77\%$ of the orbits of figure \ref{fig:3D_scan} are
characterized as chaotic, having $\mbox{GALI}_3 < 10^{-8}$. In
order to have the same percentage of orbits identified as chaotic
using SALI (i.~e.~having $\mbox{SALI} < 10^{-8}$) the same
experiment has to be carried out for $t=2000$ units, requiring
$53\%$ more CPU time. Due to the high percentage of chaotic
orbits, in this case, even when the SALI is computed for $t=1000$
the corresponding CPU time is $12\%$ higher than the one needed
for the computation of figure \ref{fig:3D_scan}, while only $55\%$
of the orbits are identified as chaotic. Thus it becomes evident
that a carefully designed application of GALI$_3$ -- or GALI$_k$
for that matter -- can significantly diminish the computational
time needed for a reliable discrimination between regions of order
and chaos in Hamiltonian systems with $N>2$ degrees of freedom.

\subsection{A multi--dimensional Hamiltonian system}
\label{App_15D}

Let us finally turn to a much higher--dimensional Hamiltonian
system having 15 degrees of freedom, i.~e.~the one shown in
(\ref{eq:15DHam}). With fixed boundary conditions
\begin{equation}
q_0(t)=q_{16}(t)=0,\,\,\, \forall t, \label{eq:boundary_con}
\end{equation}
it is known that there exists, for all energies, $H_{15}=E$, a
simple periodic orbit, satisfying \cite{OHS69,ABS05}
\begin{equation}
q_{2i}(t)=0,\,\,\,q_{2i-1}(t)=-q_{2i+1}(t)=q(t),\,\,\,
i=1,2,\ldots,7, \label{eq:15D_initials}
\end{equation}
where $q(t)=q(t+T)$ obeys a simple nonlinear equation admitting
Jacobi elliptic function solutions. For the parameter values
$H_{15}=26.68777$ and $\beta=1.04$ used in an earlier study
\cite{ABS05}, we know that this orbit is unstable and has a
sizable chaotic region around it. As initial conditions for
(\ref{eq:15D_initials}) we take
\begin{equation}
q(0)=1.322 \,\,\, \mbox{and} \,\,\, p_i(0)=0, \,\,\,
i=1,2,\ldots,15. \label{eq:15D_incon}
\end{equation}

\begin{figure}
\centerline{\includegraphics[width=7.5 cm,height=7.5 cm]{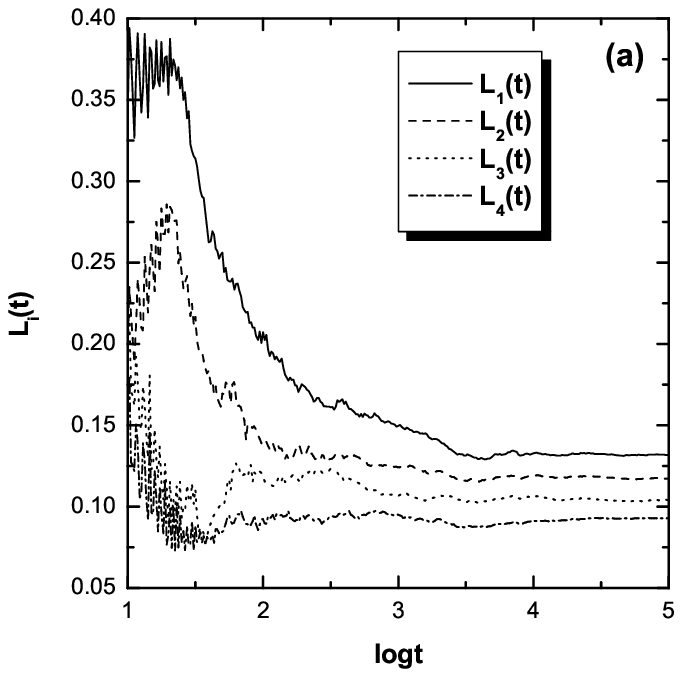}
\includegraphics[width=7.4 cm,height=7.4 cm] {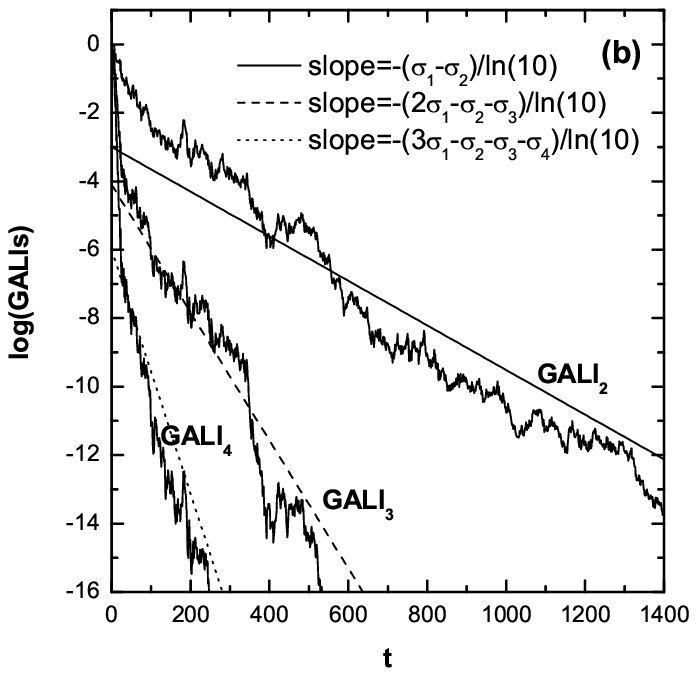} }
\caption{(a) The evolution of $L_1(t)$, $L_2(t)$, $L_3(t)$ and
$L_4(t)$ for a chaotic orbit  of the 15D system (\ref{eq:15DHam}).
(b) The evolution of GALI$_2$, GALI$_3$ and GALI$_4$ for the same
orbit. The plotted lines correspond to functions proportional to
$e^{-(\sigma_1 - \sigma_2)t}$, $e^{-(2\sigma_1 - \sigma_2-
\sigma_3)t}$ and $e^{-(3\sigma_1 - \sigma_2- \sigma_3-
\sigma_4)t}$,  for $\sigma_1=0.132$, $\sigma_2=0.117$,
$\sigma_3=0.104$, $\sigma_4=0.093$. Note that the $t$--axis is
linear.} \label{eq:15DHam_ch_GALIs}
\end{figure}

First, we consider a chaotic orbit which is located close to this
periodic solution, by taking as initial conditions $q_1(0)=q(0)$,
$q_3(0)=q_7(0)=q_{11}(0)=-q(0)+10^{-7}$,
$q_5(0)=q_9(0)=q_{15}(0)=q(0)-10^{-7}$, $q_{2i}=0$ for
$i=1,2,\ldots,7$ and $p_i(0)=0$ for $i=1,2,\ldots,14$,
$p_{15}(0)=0.00323$. The chaotic nature of this orbit is revealed
by the fact that its maximal LCE is positive (see figure
\ref{eq:15DHam_ch_GALIs}(a)).
In fact, from the results of figure \ref{eq:15DHam_ch_GALIs}(a) we
deduce reliable estimates of the system's four largest Lyapunov
exponents: $\sigma_1 \approx 0.132$, $\sigma_2 \approx 0.117$,
$\sigma_3 \approx 0.104$ and $\sigma_4 \approx 0.093$. Thus, we
have a case where several LCEs have positive values, the largest
two of them being very close to each other. The behavior of the
GALIs is again quite accurately approximated by the theoretically
predicted exponential laws (\ref{eq:ch_GALI_decay}). This becomes
evident by the results presented in figure
\ref{eq:15DHam_ch_GALIs}(b), where we plot the time evolution of
GALI$_2$, GALI$_3$ and GALI$_4$ as well as the exponential laws
that theoretically describe the evolution of these indices. In
this case, GALI$_2$ does decay to zero relatively slowly since
$\sigma_1$ and $\sigma_2$ have similar values and hence, using
GALI$_3$, GALI$_4$ or a GALI of higher order, one can determine
the chaotic nature of the orbit much faster.

\begin{figure}
\centerline{\includegraphics[width=7.5 cm,height=7.5 cm]{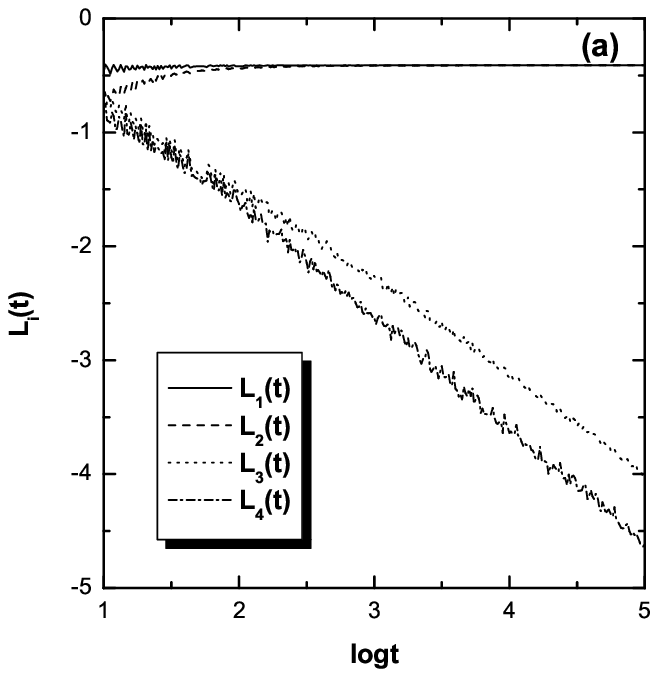}
\includegraphics[width=7.5 cm,height=7.5 cm] {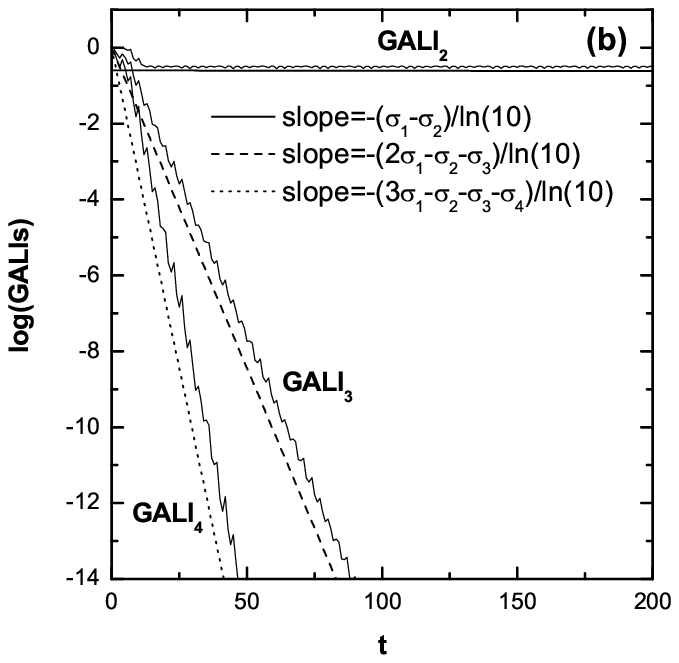} }
\caption{(a) The evolution of $L_1(t)$, $L_2(t)$, $L_3(t)$ and
$L_4(t)$ for an orbit which is very close to the unstable periodic
orbit (\ref{eq:15D_incon}) of the 15D system (\ref{eq:15DHam}).
(b) The evolution of GALI$_2$, GALI$_3$ and GALI$_4$  of the same
orbit. The plotted lines correspond to functions proportional to
$e^{-(\sigma_1 - \sigma_2)t}$, $e^{-(2\sigma_1 - \sigma_2-
\sigma_3)t}$ and $e^{-(3\sigma_1 - \sigma_2- \sigma_3-
\sigma_4)t}$,  for $\sigma_1=0.3885$, $\sigma_2=0.3883$,
$\sigma_3=0$, $\sigma_4=0$. Note that the $t$--axis is linear.}
\label{eq:15DHam_ch_GALIs_po}
\end{figure}

It is worth mentioning that (\ref{eq:ch_GALI_decay}) describes
much more accurately the evolution of GALI$_k$  when the  orbit we wish
to study is very close to the unstable periodic solution
(\ref{eq:15D_initials}) itself. This is due to the fact
that in that case, the LCEs are directly  related to the eigenvalues of the monodromy matrix  associated with
the variational equations of this unstable periodic orbit, see equation (\ref{eq:sigma_lambda}). In
fact, for our choice of parameters, this matrix has two equal
pairs of real eigenvalues with magnitude greater than one, while
all other eigenvalues lie on the unit circle in the complex plane.
As a consequence, the orbit has two nearly identical positive
Lyapunov exponents (as well as their two negative counterparts),
while all other exponents are zero. This is shown in figure
\ref{eq:15DHam_ch_GALIs_po}(a),
where we plot the evolution of the $L_i(t)$ for $i=1,2,3,4$, whose
limits for $t\rightarrow \infty$ are the 4 largest Lyapunov
exponents. From these results we deduce  $\sigma_1 \approx
0.3885$, $\sigma_2 \approx 0.3883$, while the decrease of $L_3(t)$
and $L_3(t)$  to zero indicate that $\sigma_3=\sigma_4=0$. In
figure \ref{eq:15DHam_ch_GALIs_po}(b) we now observe that GALI$_2$
remains practically constant for this particular time interval
(actually it decreases to zero extremely slowly following the
exponential law $e^{-(\sigma_1-\sigma_2)t}=e^{-0.0002t}$). On the
other hand, GALI$_3$ and GALI$_4$ decay exponentially to zero
following the laws, $\mbox{GALI}_3 \propto e^{-(2\sigma_1 -
\sigma_2- \sigma_3)t}$, $\mbox{GALI}_3 \propto e^{-(3\sigma_1 -
\sigma_2- \sigma_3- \sigma_4)t}$, given by equation
(\ref{eq:ch_GALI_decay}).

\section{Discussion and conclusions}
\label{Summary}

In this paper we have introduced and applied the Generalized
Alignment Indices of order $k$ (GALI$_k$) as a tool for studying
local and global dynamics in conservative dynamical systems, such
as Hamiltonian systems of $N$ degrees of freedom, or $2N$--dimensional
symplectic maps. We have shown that these indices can be
successfully employed not only to distinguish individual orbits as
chaotic or regular, but also to efficiently chart large domains of
phase space, characterizing the dynamics in the various regions by
different behaviors of the indices ranging from regular (GALI$_k$s
are constant or decay by well--defined power laws) to chaotic
(GALI$_k$s exponentially go to zero).

A different approach than simply calculating the maximal Lyapunov
exponent is to compute the so--called Smaller Alignment Index
(SALI), following the evolution of {\it two} initially different
deviation vectors. This approach has been used by several authors
and has proved quite successful, as it can determine the nature of
the dynamics more rapidly, reliably and efficiently than the
maximal LCE. In the present paper, motivated by the observation
that the SALI is in fact proportional to the `area' of a
parallelogram, having as edges the two normalized deviation
vectors, we have generalized SALI by defining a quantity called
GALI$_k$, representing the `volume' of a parallelepiped having as
edges $k>2$ initially linearly independent unit deviation vectors. In
practice, GALI$_k$ is computed as the `norm' of the `exterior' or
wedge product of the $k$ normalized deviation vectors.

For the numerical evaluation of GALI$_k$, we need to compute the
reference orbit we are interested in from the fully nonlinear
equations of the system, as well as follow the time evolution of
$k$ deviation vectors, solving the (linear) variational equations
about the orbit. How many such vectors should we take? Since the
phase space of the dynamical system is $2N$--dimensional, $k$
should be less than or equal to $2N$, otherwise GALI$_k$ will be equal to zero
already from the start. However, even though we may choose our
deviation vectors initially linearly independent, they may {\it
become dependent} as time evolves, in which case the phase space
`volume' represented by GALI$_k$ will vanish! This is precisely
what happens for all $k>2$ if our reference orbit is chaotic, and
also if it is regular and $k>N$, but at very different time rates.

In particular, we showed analytically and verified numerically in
a number of examples of Hamiltonian systems that for chaotic
orbits GALI$_k$ tends exponentially to zero following a rate which
depends on the values of several Lyapunov exponents (see equation
(\ref{eq:ch_GALI_decay})). On the other hand, in the case of
regular orbits, GALI$_k$ with $2\leq k \leq N$ fluctuates around
non--zero values, while, for $N < k \leq 2N$, it tends to zero
following a power law (see equation (\ref{eq:GALI_order_all})).
The exponent of the power law depends on the values of $k$ and
$N$, as well as on the number $m$ of deviation vectors that may
have been chosen initially tangent to the torus on which the orbit
lies.

Clearly, these different behaviors of the GALI$_k$ can be
exploited for the rapid and accurate determination of the chaotic
versus regular nature of a given orbit, or of an ensemble of
orbits. Varying the number of deviation vectors (and bringing more
LCEs into play), we can, in fact, achieve high rates of
identification of chaotic regions, in a computationally
advantageous way. Secondly, regular motion can be identified by
the index being nearly constant for small $k$, while, when $k$
exceeds the dimension of the orbits' subspace, GALI$_k$ decays by
well--defined power laws. This may help us identify, for example,
cases where the motion occurs on cantori of dimension $d<N$ (see
e.g. \cite{LL92}) and the orbits  become `sticky' on island
chains, before turning truly chaotic and exponential decay takes
over.

We have also studied on specific Hamiltonians with $N>2$ the {\it
computational efficiency} of the GALI$_k$. One might suspect, of
course, that the best choice would be GALI$_N$ since this is the
index that exhibits the most different behavior for regular and
chaotic orbits. On the other hand, it is clear that following a
great number of deviation vectors requires considerably more
computation time. It turns out, however, that, if chaos occupies a
`large' portion of phase space, a well--tailored application of
GALI$_k$, with $2<k\leq N$, can significantly diminish the CPU
time required for the detailed `charting' of phase space, compared
with SALI ($k=2$), as we demonstrated on specific examples in
section \ref{App_3D} (see figure \ref{fig:3D_scan}).

Although the results presented in this paper were obtained for $N$
degree of freedom Hamiltonian systems, it is easy to see that they
also apply to $2N$--dimensional symplectic maps. So, equations
(\ref{eq:ch_GALI_decay}) and (\ref{eq:GALI_order_all}) which
describe the behavior of GALI$_k$, with $2 \leq k \leq 2N$, for
chaotic and regular orbits respectively are expected to hold in
that case also. One remark is in order, however: In the case of
$N=1$, i.~e.~for 2D maps, the first condition of equation
(\ref{eq:GALI_order_all}) cannot be fulfilled. Thus, for regular
orbits of 2D maps, any 2 initially independent deviation vectors
will become aligned in the direction tangent to the corresponding
invariant curve and GALI$_2$ will tend to zero following a power
law of the form $\mbox{GALI}_2 \propto 1/t^2$. This behavior is
already known in the literature \cite{S01}.

\ack

This work was partially supported by the European Social Fund
(ESF), Operational Program for Educational and Vocational Training
II (EPEAEK II) and particularly the Programs HERAKLEITOS,
providing a Ph.~D.~scholarship for the third author (C.~A.) and
the Program PYTHAGORAS II, partially supporting the first author
(Ch.~S.). Ch.~S. was also supported by the Marie Curie
Intra--European Fellowship No MEIF--CT--2006--025678. The second
author (T.~B.) wishes to express his gratitude to the beautiful
Centro Internacional de Ciencias of the Universidad Autonoma de
Mexico for its excellent hospitality during his visit in January
-- February $2006$, when some of this work was completed. In
particular, T.~B.~wants to thank the main researchers of this
Center, Dr.~Christof Jung and Thomas Seligman for numerous
conversations on the stability of multi--dimensional Hamiltonian
systems. Finally, we would  like to thank the referees for very
useful comments which helped us improve the clarity of the paper.

\appendix

\section{Wedge product}
\label{Wedge}

Following an introduction to the theory of wedge products as
presented in textbooks such as \cite{Spivak_1999}, let us consider
an $M$--dimensional vector space $V$ over the field of real
numbers $\mathbb{R}$. The exterior algebra of $V$ is denoted by
$\Lambda(V)$ and its multiplication, known as the wedge product or
the exterior product, is written as $\wedge$. The wedge product is
associative:
\begin{equation}
(\vec{u} \wedge \vec{v}) \wedge \vec{w}= \vec{u} \wedge (\vec{v}
\wedge \vec{w})  \label{eq:w_a}
\end{equation}
for $\vec{u}, \vec{v}, \vec{w} \in V$ and bilinear
\begin{eqnarray}
 (c_1 \vec{u} + c_2 \vec{v}) \wedge \vec{w}= c_1(\vec{u}\wedge
\vec{w}) + c_2(\vec{v}\wedge \vec{w}), & &  \nonumber \\ \vec{w}
\wedge (c_1 \vec{u} + c_2 \vec{v}) = c_1(\vec{w}\wedge \vec{u}) +
c_2(\vec{w}\wedge \vec{v}) & &  \label{eq:w_b}
\end{eqnarray}
for $\vec{u}, \vec{v}, \vec{w} \in V$ and $c_1, c_2 \in
\mathbb{R}$. The wedge product is also alternating on $V$
\begin{equation}
\vec{u}\wedge \vec{u}=\vec{0} \label{eq:w_c}
\end{equation}
for all vectors $\vec{u}\in V$. Thus we have that
\begin{equation}
\vec{u}\wedge \vec{v}=- \vec{v}\wedge\vec{u} \label{eq:w_d}
\end{equation}
for all vectors $\vec{u},\vec{v} \in V $ and
\begin{equation}
\vec{u}_1\wedge \vec{u}_2\wedge \cdots \wedge\vec{u}_k=\vec{0}
\label{eq:w_e}
\end{equation}
whenever $\vec{u}_1, \vec{u}_2,\ldots ,\vec{u}_k\in V$ are
linearly dependent.

Elements of the form $\vec{u}_1\wedge \vec{u}_2\wedge \cdots
\wedge\vec{u}_k$ with $\vec{u}_1, \vec{u}_2,\ldots ,\vec{u}_k\in
V$ are called $k$--vectors. The subspace of $\Lambda(V)$ generated
by all $k$--vectors is called the $k$--th exterior power of $V$
and denoted by $\Lambda^k(V)$. The exterior algebra $\Lambda(V)$
can be written as the direct sum of each of the $k$--th powers of
V:
\begin{equation}
\Lambda(V)= \bigoplus_{k=0}^{M}\Lambda^k(V)= \Lambda^0(V) \oplus
\Lambda^1(V) \oplus\Lambda^1(V) \oplus \cdots \oplus \Lambda^M(V)
\label{eq:w_ext}
\end{equation}
where $\Lambda^0(V)=\mathbb{R}$ and $\Lambda^1(V)=V$.

Let $\{\hat{e}_1,\hat{e}_2,\ldots,\hat{e}_M\}$ be an orthonormal
basis of V, i.~e.~$\hat{e}_i$, $i=1,2,\ldots,M$ are linearly
independent vectors of unit magnitude and
\begin{equation}
\hat{e}_i \cdot \hat{e}_j = \delta_{ij}\label{eq:w_delta}
\end{equation}
where ($\cdot$) denotes the inner product in $V$ and
\begin{equation}
\delta_{ij}=\left\{ \begin{array}{ll}
1&\mbox{for} \,\,\, i=j\\
0&\mbox{for} \,\,\, i\neq j\\ \end{array} \right.
.\label{eq:w_kron}
\end{equation}
It can be easily seen that the set
\begin{equation}
\{\hat{e}_{i_1} \wedge \hat{e}_{i_2} \wedge \cdots \wedge
\hat{e}_{i_k}\,\, | \,\, 1 \leq i_1 < i_2 < \cdots < i_k \leq M \}
\label{eq:w_basis}
\end{equation}
is a basis of $\Lambda^k(V)$ since any wedge product of the form
$\vec{u}_1\wedge \vec{u}_2\wedge \cdots \wedge\vec{u}_k$ can be
written as a linear combination of the $k$--vectors of equation
(\ref{eq:w_basis}). This is true because every vector $\vec{u}_i$,
$i=1,2,\ldots,k$ can be written as a linear combination of the
basis vectors $\hat{e}_i$, $i=1,2,\ldots,M$ and using the
bilinearity of the wedge product this can be expanded to a linear
combination of wedge products of those basis vectors. Any wedge
product in which the same basis vector appears more than once is
zero, while any wedge product in which the basis vectors do not
appear in the proper order can be reordered, changing the sign
whenever two basis vectors change places. The dimension of
$\Lambda^k(V)$ is equal to the binomial coefficient
\begin{equation}
\mbox{dim}\Lambda^k(V)=\left(\begin{array}{c} M \\k \end{array}
\right)= \frac{M!}{k!(M-k)!}
 \label{eq:w_binom}
\end{equation}
and thus the dimension of $\Lambda(V)$ is equal to the sum of the
binomial coefficients
\begin{equation}
\mbox{dim}\Lambda(V)=\sum_{k=0}^{M}\left(\begin{array}{c} M \\k
\end{array} \right)= 2^M .
 \label{eq:w_binom2}
\end{equation}

The coefficients of a $k$--vector $\vec{u}_1\wedge \vec{u}_2\wedge
\cdots \wedge\vec{u}_k$ are the minors of the matrix that
describes the vectors $\vec{u}_i$, $i=1,2,\ldots,k$ in terms of
the basis $\hat{e}_i$, $i=1,2,\ldots,M$. Let us write these
relations in matrix form
\begin{equation}
\left[\begin{array}{c}
 \vec{u}_1 \\ \vec{u}_2 \\ \vdots \\
\vec{u}_k \end{array} \right] = \left[
\begin{array}{cccc}
u_{11} & u_{12} & \cdots & u_{1M} \\
u_{21} & u_{22} & \cdots & u_{2M} \\
\vdots & \vdots &  & \vdots \\
u_{k1} & u_{k2} & \cdots & u_{kM} \end{array} \right] \cdot
\left[\begin{array}{c}
 \hat{e}_1 \\ \hat{e}_2 \\ \vdots \\
\hat{e}_M \end{array} \right]= \textbf{C}\cdot
\left[\begin{array}{c}
 \hat{e}_1 \\ \hat{e}_2 \\ \vdots \\
\hat{e}_M \end{array} \right] \,\,\,   \label{eq:w_matrix}
\end{equation}
$\textbf{C}$ being the matrix of the coefficients of vectors
$\vec{u}_i$, $i=1,2,\ldots,k$ with respect to the orthonormal
basis $\hat{e}_i$, $i=1,2,\ldots,M$ and $u_{ij}$,
$i=1,2,\ldots,k$, $j=1,2,\ldots,M$ being real numbers. Then the
wedge product $\vec{u}_1\wedge \vec{u}_2\wedge \cdots
\wedge\vec{u}_k$ is defined by
\begin{equation}
\vec{u}_1\wedge \vec{u}_2\wedge \cdots \wedge\vec{u}_k = \sum_{1
\leq i_1 < i_2 < \cdots < i_k \leq M} \left|
\begin{array}{cccc}
u_{1 i_1} & u_{1 i_2} & \cdots & u_{1 i_k} \\
u_{2 i_1} & u_{2 i_2} & \cdots & u_{2 i_k} \\
\vdots & \vdots &  & \vdots \\
u_{k i_1} & u_{k i_2} & \cdots & u_{k i_k} \end{array} \right|
\hat{e}_{i_1}\wedge \hat{e}_{i_2} \wedge \cdots \wedge
\hat{e}_{i_k}
 \label{eq:w_wedge}
\end{equation}
where the sum is performed over all possible combinations of $k$
indices out of the $M$ total indices. So the coefficient of a
particular $k$--vector $\hat{e}_{i_1}\wedge \hat{e}_{i_2} \wedge
\cdots \wedge \hat{e}_{i_k}$ is the determinant of the $k\times k$
submatrix of the $k \times M$ matrix of coefficients appearing in
equation (\ref{eq:w_matrix}) formed by its $i_1$, $i_2$, $\ldots$,
$i_k$ columns.

\section{The relation between GALI$_2$ and SALI}
\label{GALI_2}

\begin{pp}
\label{p:1} We consider  a $2N$--dimensional vector space over the
field of real numbers $\mathbb{R}$, which has the usual Euclidean
norm and is spanned by the orthonormal basis
$\{\hat{e}_1,\hat{e}_2,\ldots,\hat{e}_{2N}\}$. We also consider
two unit vectors $\hat{w}_1$, $\hat{w}_2$ in this space so that
\begin{equation}
\hat{w}_1=\sum_{i=1}^{2N} w_{1i}\hat{e}_i\,\,\, , \,\,\,
\hat{w}_2=\sum_{i=1}^{2N} w_{2i}\hat{e}_i,
\label{eq:proposition_1}
\end{equation}
and
\begin{equation}
\sum_{i=1}^{2N} w_{1i}^2=1\,\,\, , \,\,\, \sum_{i=1}^{2N}
w_{2i}^2=1. \label{eq:proposition_2}
\end{equation}
Let us now define the 2--vector $\hat{w}_1 \wedge \hat{w}_2$ from
equation (\ref{eq:w_wedge}) and its norm from equation
(\ref{eq:norm}). Under these assumptions the following holds:
\begin{equation}
\| \hat{w}_1 \wedge \hat{w}_2 \|=
\frac{\|\hat{w}_1-\hat{w}_2\|\cdot
\|\hat{w}_1+\hat{w}_2\|}{2}\label{eq:proposition}
\end{equation}
\end{pp}
{\it Proof}. Expanding the right hand side of equation
(\ref{eq:proposition}) we have:
\begin{eqnarray}
\label{eq:rhs}  {\mathcal{A}} =\left(
\frac{\|\hat{w}_1+\hat{w}_2\|\cdot \|\hat{w}_1-\hat{w}_2\|}{2}
\right)^2= \frac{\sum_{i=1}^{2N} \left( w_{1i}-w_{2i}\right)^2
\cdot \sum_{i=1}^{2N} \left( w_{1i}+w_{2i}\right)^2}{4} = \nonumber\\
= \frac{1}{4} \cdot \left[ \left( \sum_{i=1}^{2N} w_{1i}^2 +
\sum_{i=1}^{2N} w_{2i}^2 - 2\sum_{i=1}^{2N} w_{1i} w_{2i} \right)
\cdot \left( \sum_{i=1}^{2N} w_{1i}^2 + \sum_{i=1}^{2N} w_{2i}^2 +
2\sum_{i=1}^{2N} w_{1i} w_{2i} \right) \right]
=\nonumber\\
= \left( 1- \sum_{i=1}^{2N} w_{1i} w_{2i} \right) \cdot \left( 1+
\sum_{i=1}^{2N} w_{1i} w_{2i} \right) = 1 -
\left(\sum_{i=1}^{2N} w_{1i} w_{2i} \right)^2 \Rightarrow \nonumber\\
{\mathcal{A}} = 1 - \left(  \sum_{i=1}^{2N} w_{1i}^2 w_{2i}^2 + 2
\sum_{i<j}w_{1i} w_{2i} w_{1j} w_{2j}\right),
\,\,\,\,\,\,\,\,\,\,\,\,\,
\end{eqnarray}
where we made use of (\ref{eq:proposition_2}). On the other hand,
using equation (\ref{eq:norm}) we get for the left hand side of
equation (\ref{eq:proposition}):
\begin{eqnarray}
\label{eq:lhs_1} {\mathcal{B}}=\| \hat{w}_1 \wedge \hat{w}_2 \|^2=
\sum_{i<j} \left|
\begin{array}{cc}
w_{1 i} & w_{1 j}  \\
w_{2 i} & w_{2 j}  \end{array} \right|^2 =
\sum_{i<j}(w_{1i}w_{2j}-w_{1j}w_{2i})^2\Rightarrow \nonumber\\
{\mathcal{B}}=\sum_{i<j}w_{1i}^2 w_{2j}^2 + \sum_{i<j}w_{1j}^2
w_{2i}^2-2\sum_{i<j}w_{1i} w_{2i} w_{1j} w_{2j}  .
\end{eqnarray}
The first two sums of equation (\ref{eq:lhs_1}) contain all the
possible products of the coordinates of the two vectors except the
ones corresponding to equal indices, $i=j$.  So the quantity
${\mathcal{B}}$ can be written as follows:
\begin{eqnarray}
\label{eq:lhs_2} {\mathcal{B}}=\sum_{i\neq j}w_{1i}^2
w_{2j}^2-2\sum_{i<j}w_{1i} w_{2i} w_{1j} w_{2j} = \nonumber\\
=\sum_{i\neq j}w_{1i}^2 w_{2j}^2 + \sum_{i=1}^{2N} w_{1i}^2
w_{2i}^2 - \left( \sum_{i=1}^{2N} w_{1i}^2 w_{2i}^2
+2\sum_{i<j}w_{1i} w_{2i} w_{1j} w_{2j} \right) .
\end{eqnarray}
Now, the first two sums contain all the possible products between
the coordinates of the two vectors and so ${\mathcal{B}}$ takes
the form:
\begin{eqnarray}
\label{eq:lhs_final} {\mathcal{B}} =\sum_{i=1}^{2N}
\sum_{j=1}^{2N} w_{1i}^2 w_{2j}^2 - \left( \sum_{i=1}^{2N}
w_{1i}^2 w_{2i}^2 +2\sum_{i<j}w_{1i} w_{2i} w_{1j} w_{2j} \right)
=\nonumber\\
=\sum_{i=1}^{2N} w_{1i}^2 \cdot \sum_{i=1}^{2N} w_{2i}^2- \left(
\sum_{i=1}^{2N} w_{1i}^2 w_{2i}^2 +2\sum_{i<j}w_{1i} w_{2i} w_{1j}
w_{2j} \right)
\Rightarrow \nonumber\\
{\mathcal{B}} = 1 - \left(  \sum_{i=1}^{2N} w_{1i}^2 w_{2i}^2 + 2
\sum_{i<j}w_{1i} w_{2i} w_{1j} w_{2j}\right),
\end{eqnarray}
where we used again (\ref{eq:proposition_2}). Comparing equations
(\ref{eq:rhs}) and (\ref{eq:lhs_final}) we see that both sides of
equation (\ref{eq:proposition}) are equal and so the proof of
proposition \ref{p:1} is complete. $\blacksquare$

Using equation (\ref{eq:proposition}) as well as the definitions
of SALI (\ref{eq:SALI}) and GALI$_2$ (\ref{eq:GALI}) we conclude
that the precise relation between the two indices is
\begin{equation}
\mbox{GALI}_2=\mbox{SALI} \cdot \frac{\max \left\{ \left\|
\hat{w}_1 +\hat{w}_2  \right\| , \left\| \hat{w}_1 -\hat{w}_2
\right\| \right\}}{2}. \label{eq:SALI_GALI2}
\end{equation}
So, the two indices are proportional to each other
\begin{equation}
\mbox{GALI}_2 \propto \mbox{SALI},  \label{eq:SALI_prop GALI2}
\end{equation}
since the quantity $m=\max \left\{ \left\| \hat{w}_1 +\hat{w}_2
\right\| , \left\| \hat{w}_1 -\hat{w}_2 \right\| \right\}$ lies in
the interval $m\in [\sqrt{2},2]$. In particular, in the case of
chaotic orbits $m\rightarrow 2$ as $\mbox{SALI}\rightarrow 0$ and
eventually GALI$_2$ also vanishes, while in the case of regular
motion $m$ fluctuates around non--zero values in the above
interval $[\sqrt{2},2)$.

From the above discussion we conclude that SALI is essentially
equivalent to GALI$_2$. In practice, however, since the
computation of GALI$_2$ according to equation (\ref{eq:norm}) for
$k=2$, requires the evaluation of several $2\times 2$
determinants, it is more convenient to compute SALI in its place,
by performing the simpler computation of equation (\ref{eq:SALI}).




\end{document}